\documentclass[pre,twocolumn,aps,eqsecnum]{revtex4}
\usepackage{epsfig}

\begin{document}
\begin{titlepage}

\title{Statistical Thermodynamics of Dislocations in Solids} 
\author{J.S. Langer}
\affiliation{Kavli Institute for Theoretical Physics, Kohn Hall, University of California, Santa Barbara, CA  93106-9530}
\date{\today}
\begin{abstract}
This review is a summary of the thermodynamic dislocation theory with special emphasis, first, on the role of an effective temperature in these driven, nonequilibrium situations and, second, on the controlling dynamics of dislocation entanglements.  Materials scientists, for decades, have asserted that statistical thermodynamics is not applicable to dislocations, and that only non-predictive phenomenological methods are available for describing their behaviors.   By use of simple, first-principles analyses and comparisons with experimental data, I argue that these scientists have been wrong, and that this venerable field urgently needs to be revitalized because of its wide-ranging fundamental and technological importance. In addition to describing recent progress in understanding strain hardening, yielding, shear banding, and the like, I show that the thermodynamic dislocation theory is now leading to a first-principles understanding of brittle and ductile fracture in crystalline solids. 
\end{abstract}

\maketitle

\end{titlepage}

\section{Introduction}
\label{Intro}

Our understanding of the mechanical behaviors of crystalline solids has been based, for almost a century, on dislocation theory. In addressing this class of  nonequilibrium phenomena, materials scientists have been trying to solve some of the most fundamental  problems in statistical physics and, at the same time, trying to provide guidance for engineering applications.  Unfortunately, much of the dislocation theory that has emerged from these efforts depends on questionable and often demonstrably incorrect, phenomenological  assumptions.  As a result, this field urgently needs to be revitalized.  I argue that the thermodynamic dislocation theory (TDT) \cite{LBL-10} is a good start in this direction.  

Most of this review consists of simple descriptions of the TDT -- especially the effective temperature and the dynamics of dislocation entanglements -- based on little more than the second law of thermodynamics and dimensional analysis.  I begin in Section \ref{History} with a brief history of dislocation theory, starting with the early contributions of G.I. Taylor and others, and then focussing on the highly influential but controversial ideas of A. Cottrell. 

Sections \ref{TDT} and \ref{EOM} discuss the effective-temperature analysis and the role of dislocation entanglements.  Sections \ref{Expt} - \ref{fracture} describe experimental predictions of the TDT, ending with a brief summary of recent developments in fracture theory.  Section \ref{next} contains a list of new research opportunities opened by the developments described in the body of this paper.

\section{History}
\label{History}

It was discovered in the 1930's, notably by G.I. Taylor \cite{TAYLOR-34}, E. Orowan and colleagues, that the deformation of a crystalline solid can occur at the cost of relatively little energy via the motion of line defects that we call ``dislocations.'' These lines are the edges of partial planes of atoms. When they are driven to move across a crystalline material, the system undergoes irreversible shear deformation.   

The decades following this insight were devoted largely to studying the properties of individual dislocations, e.g. the  stress needed to move a dislocation through the lattice and the interactions between the dislocations and various crystalline defects.  Standard references from this era include \cite{COTTRELL-53,FRIEDEL-67,HIRTH-LOTHE-68}.

The major challenge was to solve the problem of work hardening, which is the question of why the stress required for deformation increases as the material is deformed. With the advent of modern microscopy, it became clear that the density of dislocations increases with deformation, so that the dislocation lines become entangled and increasingly immobile.   

In his seminal book on dislocation theory \cite{COTTRELL-53}, published in 1953, Cottrell computed the entropy of dislocations and found -- correctly -- that it is extremely small compared to the entropy of the solid as a whole.  From this he concluded -- incorrectly -- that entropic considerations are not relevant to dislocations.  Fifty years later, he wrote a preface to a volume of essays celebrating the anniversary of the publication of his book.\cite{COTTRELL-02} Here are some passages from that preface.

The opening sentence is: ``It is sometimes said that the turbulent flow of fluids is the most difficult remaining problem in classical physics.  Not so.  Work hardening is worse.''   Then, in the next paragraph, he says that ``--- neither of the two main strategies of theoretical many-body physics -- the statistical mechanical approach; and the reduction of the many-body problem to that of the behaviour of a single element of the assembly -- is available to work hardening.  The first fails because the behaviour of the whole system is governed by that of weakest links, not the average, and is thermodynamically irreversible.  The second fails because dislocations are flexible lines, interlinked and entangled, so that the entire system behaves more like a single object of extreme structural complexity and deformability, that Nabarro and I once compared to a bird's nest ---.''  Then, at the end of this preface, fifty years after publication of his book, he stated that ``... the theory is ... still at the stage of merely being interpretative, not predictive.''

Although Cottrell may have been fundamentally wrong in his understanding of nonequilibrium statistical mechanics, these remarks show that he was deeply perceptive about the physics of dislocations and the state of this research field.  Much of what follows in this review is my explanation of how we can address the thermodynamic issues correctly.  But I think that Cottrell understood quite well the main physical considerations  -- much better than his colleagues who still seem to have misunderstood them in 2021.  His bird's-nest analogy means that the many-body theorist's concept of weakly interacting quasiparticles does not work here.  Dislocations are not analogous to those quasiparticles; and strain hardening cannot be predicted simply by assigning drag forces or interactions with crystalline heterogeneities to the motions of single dislocations. Yet that is precisely what the majority of materials theorists have been trying to do for decades.  For just a few examples, see \cite{GRAY-12,ARMSTRONG-HP14,ARMSTRONG-HP16,Barton etal-2011}.

The rigidity of a real bird's nest is determined by the strength of the contact  interactions between pairs of straws.  Similarly, the rigidity of a deforming crystal must be determined by the strength of the pinning interactions that connect pairs of entangled dislocations.  Those long-lasting entanglements, and the slow, thermally activated rates at which they come apart under stress, must control the rates of strain hardening. Faster processes such as those by which the dislocation network forms and adjusts itself are irrelevant over a wide range of experimental conditions. This significance of Cottrell's bird's nest analogy seems seldom to have been appreciated.   As a result, theoretical research in this field has been -- in Cottrell's language -- ``not predictive''  since the 1950's.  

By calling this research ``not predictive,'' I certainly do not mean that nothing has  been accomplished.  On the contrary, there has been a large amount of increasingly sophisticated experimental observation of dislocations moving in complex environments.  We have seen images of dislocations interacting with other dislocations, piling up at grain boundaries, forming cellular structures, being emitted at crack tips, etc..  We also have a growing wealth of information about relations between stress, strain, strain rate, temperature, crystalline orientation, grain size, sample preparation, and the like.  The trouble is that, without basic understanding, we cannot know how these observations might be related to one another or be relevant to large-scale strength of materials. 

There also have been numerical simulations.  Much of this work uses the ``discrete dislocation dynamics'' (DDD) method in which dislocations are represented by line segments. They are produced computationally at isolated heterogeneities such as Frank-Read sources, and are driven by applied stresses to move and interact with impediments according to hypothesized dynamical rules  (e.g. see  \cite{LESARetal-06,KUBIN-08,LESAR-14,SILLS-18}).  Unfortunately, the DDD simulations generally go only to very small strains, of order 0.01 or less, whereas the interesting behaviors occur experimentally at strains of order unity.  This inability of DDD to go beyond the earliest stage of strain hardening is almost certainly caused by the fact that these simulations do not include realistic pinning interactions between the dislocation lines, and therefore are not realistic descriptions of the entanglement effects that Cottrell thought were so important.  

The situation with regard to numerical simulations is now changing with the advent of  realistic molecular dynamics simulations of three-dimensional crystals subject to shear stresses.  I am thinking specifically of recent results of Zepeda-Ruiz {\it et al} \cite{BULATOV-17}, who have used one of the world's most powerful computers at Lawrence Livermore National Laboratory. Their results are highly relevant to what I want to say here.  I shall return to them in Section \ref{LMD}.  

\section{Thermodynamic Dislocation Theory} 
\label{TDT}

My plan for what follows is to present a summary of the thermodynamic dislocation theory (TDT), focussing on its main features and describing some of the results. I even shall simplify some of the notation in hopes of clarifying the physical concepts.  More details can be found in the published literature, especially  \cite{LBL-10,JSL-17,JSL-17a,JSL-17rev,LTL-17,LTL-18,JSL-18,JSL-PCH-19}.  

The version of the TDT presented here is a caricature of the ``real'' dislocation theories described, for example, in Refs. \cite{COTTRELL-53,FRIEDEL-67,HIRTH-LOTHE-68}. My TDT dislocations are simply lines.  I do not ask whether they are edge dislocations or screw dislocations, or whether they are excess dislocations or geometrically necessary ones.  Their Burgers vectors have lengths but not directions.  The crystals through which they move might be fcc, bcc, hcp, or something else.  Their motions are unaffected by crystalline orientations or slip planes or stacking faults.  They do not undergo cross slip.  They interact  with each other only at the junctions where they are pinned and not via long-ranged elastic forces.  Also, I do not make the common  distinction between ``mobile'' and ``immobile'' dislocations, in this case because I consider that distinction to be wrong.  

All of these omissions might be restored within the framework of the TDT; but my main message is that we can go remarkably far with only this TDT caricature.  Once we see what important physics is missing, we should be able to put the realistic features back into the theory in fundamentally consistent ways and thereby understand what roles they play and how important those roles may be.  I shall be more specific about some opportunities for further investigation at the end of this review.

There are two main features that distinguish the TDT from earlier descriptions of dislocation dynamics.  These are, first and most fundamentally, the use of an effective disorder temperature for describing nonequilibrium behavior consistent with the second law of thermodynamics.  Second is a simple depinning analysis that describes the behavior of Cottrell's birds' nest and identifies the thermal activation energy.  I start with the effective temperature because it is essential to everything that follows.

\subsection{Effective Temperature}
\label{Teff}

In contrast to amorphous plasticity (e.g. see \cite{FL-11}), where identifying shear transformation zones or the like has always been problematic, the elementary flow defects in crystals -- the dislocations -- are unambiguous.  They are easily identifiable line defects, whose dynamic time scales are longer than those of the ambient thermal fluctuations by many orders of magnitude.  They have well defined energies and easily visible configurations.  As emphasized  above, they are intrinsically nonequilibrium entities in these moving systems.  They are the agents of deformation and dissipation when external forces drive energy to flow through the system.  Under the influence of these forces, the dislocations undergo complex chaotic motions, so that it becomes both possible and necessary to describe their behavior statistically.  That statistical analysis is literally thermodynamic.  

For clarity, let me state from the beginning that I shall not consider the apparently deterministic, sometimes chaotic motions of dislocations, such as those that produce acoustic emission or irregular shear banding.  Those phenomena are seen at extremely small strain rates and short times, where the systems are non-ergodic and the TDT statistical analysis ceases to be relevant.  The distinction between these two, qualitatively different kinds of behaviors is interesting; but it is not the principal focus of this review.      

The central idea here is that the dislocations and the rest of the system -- say, the fast, kinetic-vibrational modes -- constitute two weakly coupled subsets of dynamic degrees of freedom. I make the strong assumption that each of these subsystems is ergodic.  We can think of them almost as if they were two different objects, one hot and the other cold, connected by weak thermal contacts that conduct heat slowly from one  to the other.  

To explore the subsystem of dislocations, it is useful to start by thinking of a slab of material lying in the plane of an applied shear stress, undergoing only uniform (on the average), steady-state shear deformation.  Then focus only on the dislocations.  The dislocation lines oriented perpendicular to this plane are driven by the  stress to move through the system, producing shear flow.   

Let the area of this slab be $A_0$ and, for the sake of argument, let its thickness be a characteristic dislocation length, say $L_0$.  Denote the configurational energy and entropy of the dislocations in this slab by $U_0(\rho)$ and $S_0(\rho)$ respectively, where $\rho$ is the areal density of dislocations or, equivalently, the total length of dislocation lines per unit volume. The entropy $S_0(\rho)$ is computed by counting the number of arrangements of dislocations at fixed values of $U_0$ and $\rho$.

The fact that the dislocations move chaotically on deformation time scales means that they explore  statistically significant parts of their configuration space.  According to Gibbs, this configurational subsystem always moves toward states of higher, and never lower, probability.  It does this at a value of the energy $U_0$ that is determined by the balance between the input power and the rate at which energy is dissipated into a thermal reservoir (i.e. the kinetic-vibrational modes).  The method of Lagrange multipliers tells us to find this most probable state by maximizing the function $S_0 - (1/\chi)\, U_0$, and then finding the value of the multiplier $1/\chi$ for which $U_0$ has the desired value.  Thus $ \chi \equiv k_B T_{e\!f\!f}$; and the free energy to be minimized is
\begin{equation}
\label{Fdef}
F_0 = U_0 - \chi\,S_0. 
\end{equation}

Minimizing $F_0$ in Eq.~(\ref{Fdef}) determines the steady-state dislocation density, say  $\rho_{ss}$, as a function of the steady-state effective temperature $\chi_{ss}$. In the simplest approximation, $U_0 = A_0\,e_D\, \rho$, where $e_D$ is a characteristic energy of a dislocation of length $L_0$.  Similarly, we can estimate the $\rho$ dependence of the entropy $S_0$ by dividing the  area $A_0$ into elementary squares of area $a^2$, where $a$ is the minimum spacing between noninteracting dislocations, an atomic-scale length.  Then we count the number of ways in which we can distribute $\rho\,A_0$ line-like dislocations, oriented  perpendicular to the plane, among those squares. The result has the familiar form $S_0 = -\,A_0\,\rho\,\ln(a^2\,\rho) + A_0\,\rho$.  Minimizing $F_0$ with respect to $\rho$ produces the familiar Boltzmann formula, 
\begin{equation}
\label{rhoss}
\rho_{ss} = {1\over a^2}\,e^{-\,e_D/\chi_{ss}}. 
\end{equation}
We see that an appreciable density of dislocations requires a value of $ \chi_{ss}$ that is comparable to $e_D$, which is enormously larger than the ambient thermal energy $k_B\,T$.   

Next, note that $\chi$ is a measure of the configurational disorder in the material, in direct analogy to the way in which the ambient temperature $T$ determines the intensity of low-energy fluctuations.  As such, $\chi_{ss}$ must be a function primarily of the plastic strain rate $\dot\epsilon^{pl}$, which determines the rate at which the atoms and dislocations are being caused to undergo rearrangements.  (Think of this as a ``stirring'' rate.)  If this rate is slow enough that the kinetic-vibrational subsystem has time to relax between dislocation-rearrangement events, then the steady state of disorder and, consequently, the density of dislocations should be constants as a function of this rate.

This argument means that $\chi_{ss}$ must have some constant nonzero value, say $\chi_0$, at strain rates appreciably smaller than atomic vibration frequencies; that is, roughly, $\dot\epsilon^{pl} \le 10^6/s$, which is true for all but strong-shock experiments and MD simulations. We can even make a rough estimate of $\chi_0$ by guessing (in the spirit of Lindemann's melting criterion) that the transition to a rate-dependent $\chi_{ss}$ occurs when the average spacing between dislocations is reduced to about ten times the minimum spacing $a$; i.e., from Eq.(\ref{rhoss}), $e_D/\chi_0 \sim 2\,\ln(10) \sim 4$.  The resulting value $\tilde \chi_0 \equiv \chi_0/e_D \sim 0.25$ is quite close to what is found experimentally. (Scaling effective temperatures by the dislocation energy $e_D$ lends some universality to this analysis.)

It follows from Eq.(\ref{rhoss}) that $\rho_{ss}$ is independent of strain rate under essentially all steady-state experimental conditions.  As will be seen in the next subsection, the driving stress is determined primarily by the dislocation density, and therefore must also be nearly independent of strain rate. In fact, the steady-state stress for room-temperature copper increases by less than a factor of $2$ between strain rates of $10^{-3}/s$ and $10^8/s$.  (See Figs. 6 and 7 in \cite{LBL-10}, which are based on data from \cite{PTW-03}.)  I find it remarkable that this previously unexplained major feature of the experimental data can be understood using just the concept of the effective temperature and some simple, dimensional arguments.  

This situation changes in interesting ways when we look at very high strain rates, for example, in strong-shock experiments or in the Livermore MD simulations \cite{BULATOV-17}.  I shall return to these issues.

\subsection{Depinning Mechanism}
\label{Depinning}

Now return to Cottrell's remark about the behavior of the entangled dislocation system being governed by its ``weakest links.'' My proposed solution to this birds'-nest problem is to assume that the dominant rate-controlling mechanism during deformation is thermally activated, localized depinning of the entangled dislocations.  

The depinning analysis starts with Orowan's relation between the plastic strain rate $\dot\epsilon^{pl}$, the dislocation density $\rho$, and the average dislocation velocity $v$:
\begin{equation}
\label{Orowan}
\dot\epsilon^{pl}= \rho\,b\,v.
\end{equation}
Here, $b$ is the magnitude of the Burgers vector which, for present purposes, is the shear displacement produced by a passing dislocation, a length of the order of atomic spacings. This equation is basically dimensional analysis. It must be correct up to some numerical prefactor if $b$ and $\ell =1/\sqrt{\rho}$ are the only two relevant length scales in the problem. It will not necessarily be correct if there is another relevant length scale, for example, the radius of curvature of a crack tip (see Sec. \ref{fracture}) that is comparable to $b$ and $\ell$. 

If a depinned dislocation segment moves almost instantaneously across the average distance between pinning sites of order $\ell$, then $v = \ell/\tau_P$, where $1/\tau_P$ is a thermally activated depinning rate given by
\begin{equation}
\label{tauP}
{1\over \tau_P} = {1\over \tau_0}\,e^{- U_P(\sigma)/k_B T},
\end{equation} 
and $\tau_0$ is a microscopic time scale. 
 
The activation barrier $U_P(\sigma)$ must be a decreasing function of the deviatoric stress $\sigma$. For dimensional reasons, $\sigma$ should be expressed in units of some physically relevant stress, which we can identify as the Taylor stress $\sigma_T$ for the following reason.  Suppose that a pinned pair of dislocations must be separated by a distance $a' \ll a$ in order to break the bond between them. If these dislocations remain pinned to other dislocations at average spacings of order $\ell$,  then this displacement is equivalent to a strain of order $a'/\ell= a'\sqrt{\rho}$ and a corresponding stress of order $\mu\,a'\,\sqrt{\rho}$, where $\mu$ is the shear modulus.   Thus  
\begin{equation}
\label{sigmaT}
\sigma_T (\rho)= \mu\,{a'\over \ell} \equiv \mu_T\,\sqrt{a^2\,\rho};~~~\mu_T = (a'/a)\,\mu,
\end{equation}
where $\sigma_T$ is the Taylor stress, rederived here by an argument roughly equivalent to the one that Taylor used in his 1934 paper.\cite{TAYLOR-34}  As in \cite{LBL-10}, write
\begin{equation}
\label{UP}
U_P(\sigma) = k_B\,T_P\,e^{- \sigma/\sigma_T(\rho)},
\end{equation}
where $k_B\,T_P$ is the pinning energy at zero stress. The exponential function used here has no special significance; it is just a simple decreasing function of  $\sigma/\sigma_T$ that neither vanishes nor diverges at finite values of its agument. The Orowan formula for the strain rate in Eq.(\ref{Orowan}) becomes: 
\begin{equation}
\label{qdef}
\dot\epsilon^{pl} \equiv {q\over \tau_0}= {b\over \tau_0}\,\sqrt{\rho}\, \exp \Bigl[- {T_P\over T} e^{-\sigma/\sigma_T(\rho)}\Bigr],
\end{equation}\\
where $q$ is a convenient shorthand for the dimensionless plastic strain rate.

Now solve Eq.(\ref{qdef}) for $\sigma$ as a function of $\rho$, $q$, and  $T$. The result is:
\begin{equation}
\label{sigmadef}
\sigma = \sigma_T(\rho)\,\,\nu(\rho,q,T),
\end{equation} 
where 
\begin{equation}
\label{nudef}
\nu(\rho,q,T) = \ln\Bigl({T_P\over T}\Bigr) - \ln\Biggl[\ln\Bigl({b\sqrt{\rho}\over q}\Bigr)\Biggr] .
\end{equation}
Note that $\nu$ is a very slowly varying function of its arguments, consistent with the well known but approximate validity of the Taylor formula in Eq.(\ref{sigmaT}). This result is also consistent with the observation at the end of the preceding subsection that, if $\rho$ is independent of strain rate, then the steady-state stress must also be very nearly a constant.  The converse of this observation is that the strain rate given by the double-exponential formula in Eq.(\ref{qdef}) is an extremely rapidly varying function of the stress and the temperature.  As will be seen below, this solution of the birds'-nest problem solves other long-standing puzzles about yielding transitions, banding instabilities and the like. 

\subsection{Scaling Test of the Steady-State TDT}
\label{Scaling}

These steady-state results, Eqs.(\ref{sigmadef}) and (\ref{nudef}), are highly unconventional in the dislocation-theory literature.  Before generalizing them to time-dependent behaviors such as strain hardening, it will be useful to show that they pass a stringent scaling test in comparison with experimental data.  

In \cite{KCL-20,JSL-KCL-20}, Le and I pointed out that the combination of Eqs.(\ref{sigmadef}) and (\ref{nudef}), in those steady-state situations where the dislocation density $\rho$ remains constant, can be used as a scaling relation between the dimensionless stress ratio $\sigma/\sigma_T$ and the dimensionless strain-rate $q$.  This relation involves only three system-dependent but theoretically strain-rate independent groups of parameters: $(\mu_T(T)\, a'\,\sqrt{\rho})$,  $(b\,\sqrt{\rho}\,\tau_0)$, and $T_P$.  Because $\mu_T$ is proportional to the shear modulus $\mu$, whose temperature dependent values are known independently,  the experimental data should collapse onto a single curve $\sigma/\sigma_T = \nu(q)$ when the values of these three system-dependent parameters are chosen correctly.  

Indeed, this scaling prediction is  remarkably well satisfied.  Figure 1 shows the scaling curve with experimental data for copper and aluminum obtained by Samanta \cite{SAMANTA} in 1971.  These are old results, but they have the advantage for us of using two different materials and testing them at different temperatures and strain rates under otherwise identical conditions.   Our scaling graph contains $32$ points: $12$ for pure copper at three temperatures in the range ($600\,C - 900\,C$) and four strain rates ($960\,/s -2300/\,s$), and $20$ for pure aluminum at four temperatures in the range ($250\,C - 550\,C)$ and five strain rates ($520\,/s - 2200\,/s$).  Clearly, these points fall accurately on the smooth curve predicted by the TDT analysis, which adds greatly to our confidence in this theory. The most physically interesting fitting parameters are $T_P = 45,000\,K$ for copper and $T_P = 27,800\,K$ for aluminum, which differ somewhat from previous estimates, possibly because of differing sample preparations or measurement techniques.   

\begin{figure}[h]
\begin{center}
\includegraphics[width=.8 \linewidth] {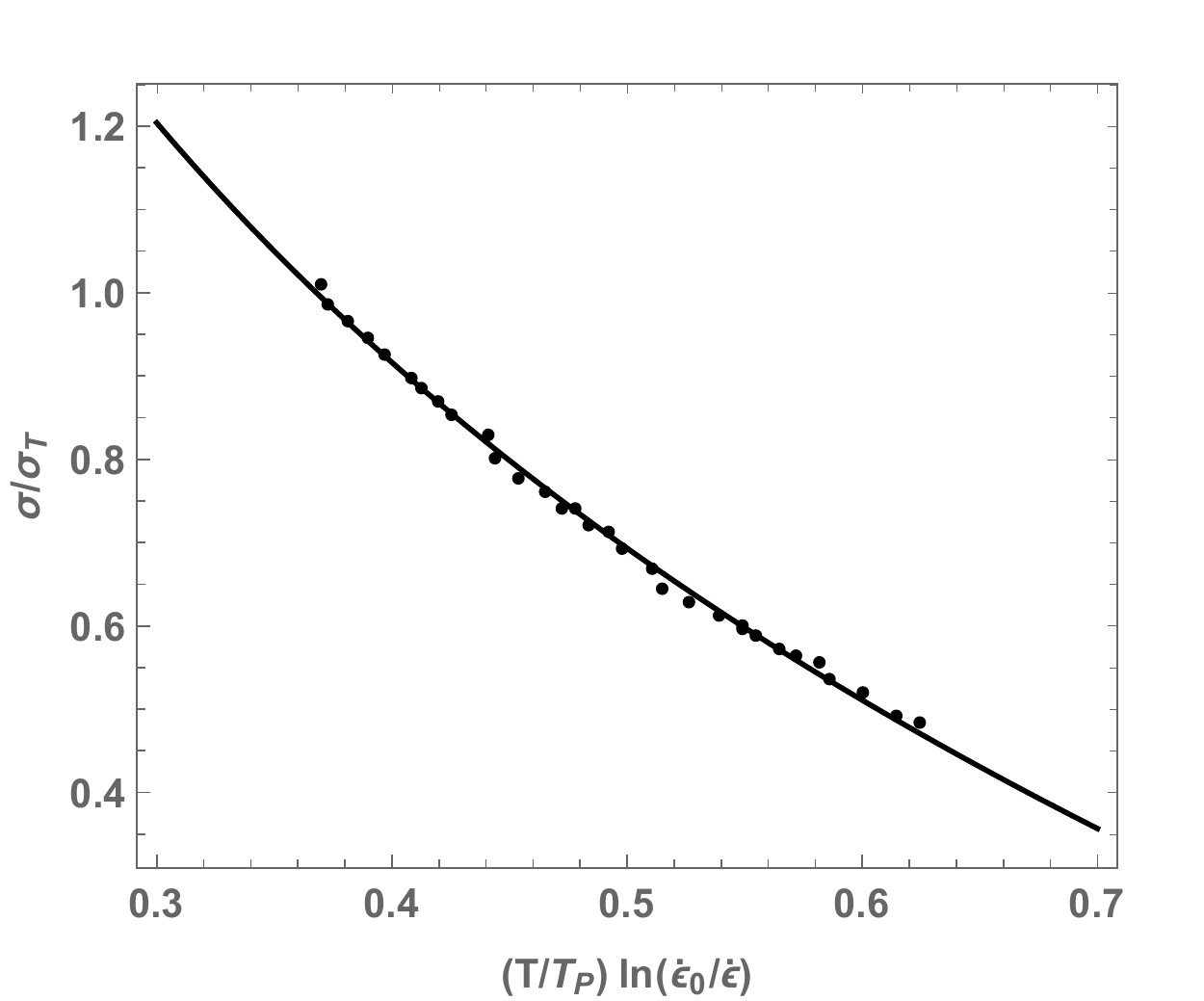}
\caption{Scaling relation implied by Eqs. (\ref{sigmadef}) and (\ref{nudef}).  The solid curve is the function $f(x) = \ln(1/x)$ with $x = (T/T_P)\,\ln(\dot\epsilon^{ss}_0/\dot\epsilon^{pl})$ and $\dot\epsilon^{ss}_0= b \sqrt{\rho}/\tau_0$.}\label{Fig1}
\end{center}
\end{figure}

In my opinion, this analysis strongly supports the fundamental principles of the TDT. It is a test of a physics-based prediction, not a phenomenological fit to data.  New experiments similar to those of Samanta, and similar scaling analyses, would be useful to explore the limits of validity of the theory. 

\section{Equations of Motion}
\label{EOM}

The discussion so far has pertained to steady-state deformations.  To address issues such as strain hardening, however, we need time dependent equations of motion.    

Start by assuming that the elastic and plastic shear rates are  additive (``hypo-elasto-plastic''), i.e. $\dot\epsilon^{total} = \dot\epsilon^{el} + \dot\epsilon^{pl}$. Then the equation of motion for the stress $\sigma$ is
\begin{equation}
\label{dotsigma}
\dot\sigma = \mu\,\dot\epsilon^{el}= \mu\,(\dot\epsilon^{total} - \dot\epsilon^{pl}),
\end{equation}
where $\mu$ is the shear modulus.  (For simplicity, I have supressed the Poisson ratio in my definition of $\mu$.) The crucial ingredient here is $\dot\epsilon^{pl}$, which is given in Eq.(\ref{qdef}) as a function of the dynamical variables $\sigma$, $\rho$, and $T$. Thus, this equation, like the others that follow, is highly nonlinear.

The equation of motion for the dislocation density $\rho$ is a statement of energy conservation:
\begin{equation}
\label{dotrho} 
\dot\rho = \kappa_{\rho}\,{\sigma\,\dot\epsilon^{pl}\over \gamma_D}\,\Bigl[1- {\rho\over \rho_{ss}(\tilde\chi)}\Bigr],
\end{equation}
where $\gamma_D \sim e_D/ L_0$ is the dislocation energy per unit length, and $\kappa_{\rho}$ is the fraction of the input power $\sigma\,\dot\epsilon^{pl}$   that is converted into dislocations.  The second term inside the square brackets determines the rate at which dislocations are annihilated.  It does this by invoking a detailed-balance approximation using the scaled effective temperature $\tilde\chi$; that is, it says that the density $\rho$ must approach the value given by Eq.(\ref{rhoss}), but with the steady-state $\tilde\chi_{ss}$ replaced by a time dependent $\tilde\chi$ during the approach to steady state deformation. 
 
Note that Eq.(\ref{dotrho}) describes the flow of energy in and out of the subsystem of dislocations.  It uses the effective temperature in an essential way; we would not have been able to write this equation without that thermodynamic basis for the theory.  With it, however, we do not need detailed information about the mechanisms by which the dislocations are annihilated.  To a first approximation, we simply need to require that those mechanisms be consistent with the second law of thermodynamics.  

All of the detailed physical ingredients of this equation are contained in the conversion factor $\kappa_{\rho}$, which describes dislocation creation.  But now the flow of information is reversed in comparison with what happened with the phenomenological curve-fitting procedures.  The general structure of Eq.(\ref{dotrho}) is not controversial; it is based on well understood physical principles.  So, by measuring the dependence of $\kappa_{\rho}$ on quantities such as strain rate or grain size or the like, we learn new physics.  

The equation of motion for $\tilde\chi$ is a statement of the first law of thermodynamics for the subsystem of dislocations, which can be written in the form 
\begin{equation}
\label{dotchi}
c_{e\!f\!f}\,e_D\,\dot{\tilde\chi} = \sigma\,\dot\epsilon^{pl}\,\Bigl( 1- {\tilde\chi\over\tilde\chi_{ss}}\Bigr) - \gamma_D\,\dot\rho.
\end{equation}\\
Here, $c_{e\!f\!f}$ is the effective specific heat, given by $V_0\,c_{e\!f\!f} = \chi\,\partial S_0/\partial\chi$, with $V_0$ being the volume and $S_0$ the dislocation entropy. The second term in the parentheses is proportional to the rate at which effective heat is converted to ordinary heat. This term assures us that $\tilde\chi$ is a thermodynamically well-defined temperature. Like the comparable term in Eq. (\ref{dotrho}), this is a detailed-balance approximation. At low to moderate strain rates, according to the discussion in Sec. \ref{Teff}, $\tilde\chi_{ss} \cong \tilde\chi_0 \sim 0.25$.  At high strain rates, $\tilde\chi_{ss}$ must be a function of $\dot\epsilon^{pl}$, as will be discussed in Sec.VI. The last term in this equation is the rate of energy storage in the form of dislocations.  It has almost always turned out to be negligible. 

Finally, because $\dot\epsilon^{pl}$ in Eq.(\ref{qdef}) is such a rapidly varying function of $T$, we need an equation of motion for the ordinary temperature.  This is 
\begin{equation}
\label{dotT}
c_T \dot T = \beta\,\sigma\,\dot\epsilon^{pl} - {\cal K}_0\,(T - T_0) + {\cal K}_1\,\nabla^2 T,
\end{equation}
where $c_T$ is the ordinary thermal specific heat, $\beta$ is the Taylor-Quinney factor that determines what fraction of the input power is converted directly into heat, $T_0$ is the ambient temperature, and ${\cal K}_0$ and ${\cal K}_1$ are thermal transport coefficients.  

\section{Comparisons with Experiment at Moderate Strain Rates}
\label{Expt}

\begin{figure}[h]
\begin{center}
\includegraphics[width=\linewidth] {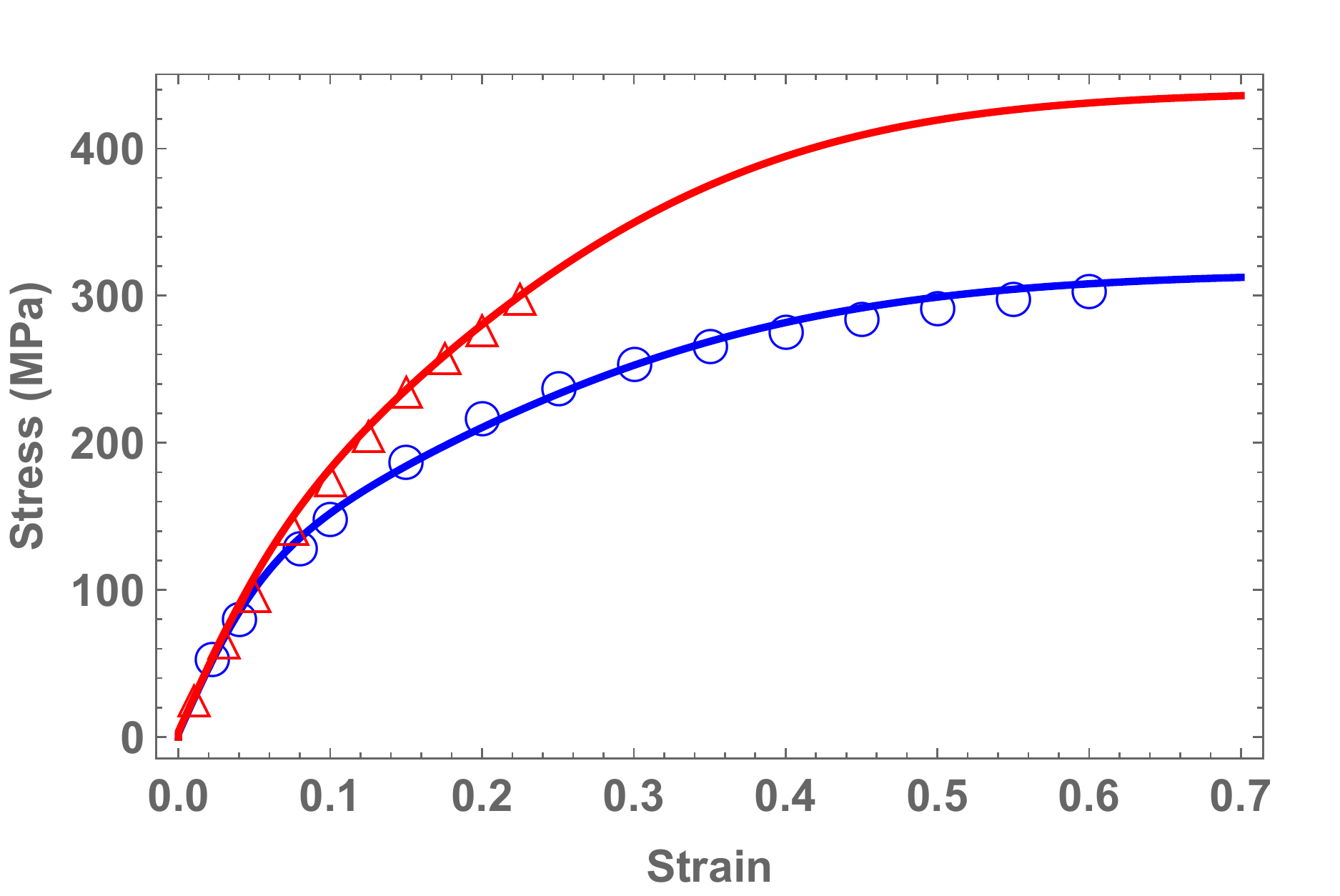}
\caption{Experimental data and theoretical stress-strain curves for copper at $T=298\,K$, for strain rates $0.002\,/s$ (lower blue curve) and $2,000\,/s$ (upper red curve).}   \label{Fig2}
 \end{center}
\end{figure}

\begin{figure}[h]
\begin{center}
\includegraphics[width=\linewidth] {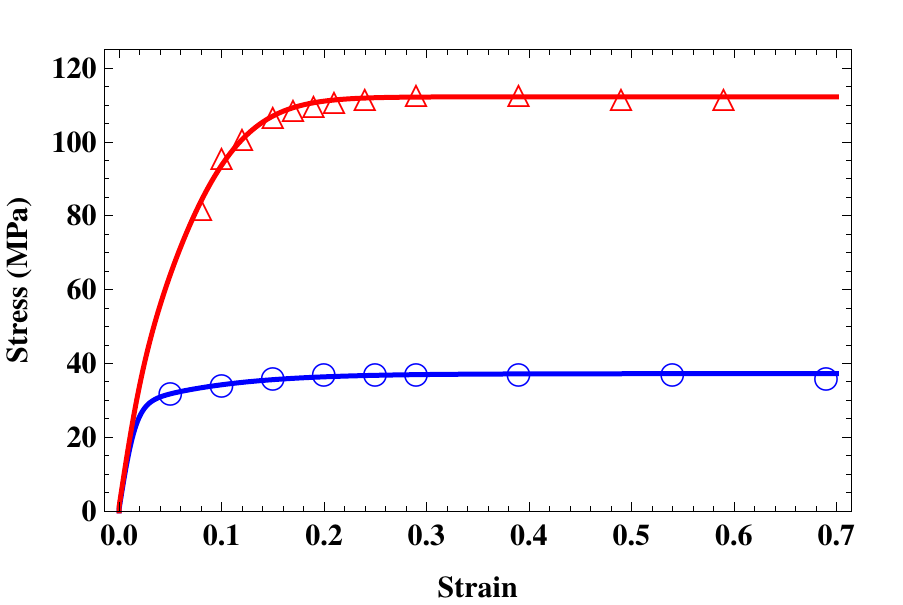}
\caption{Experimental data and theoretical stress-strain curves for copper at $T=1173\,K$, for strain rates $0.066\,/s$ (lower blue curve) and $980\,/s$ (upper red curve).}   \label{Fig3}
 \end{center}
\end{figure}

Solutions of Eqs.(\ref{dotsigma} - \ref{dotT}) and comparisons with experimental data have been published in Refs.\cite{LBL-10,JSL-17,JSL-17a,JSL-17rev,LTL-17,LTL-18,JSL-18,JSL-PCH-19}.  Here I  summarize a few of those results to illustrate points made in the preceding discussion.  More details, including all parameter values for these selected cases, can be found in the cited papers. 

{\it Strain Hardening for Copper:} \, Start with the strain-hardening curves for copper that are shown in Figs. \ref{Fig2}  and \ref{Fig3} for four widely different strain rates and temperatures.  The experimental points are taken from \cite{LANL-99}.  The four theoretical curves, taken from \cite{JSL-17rev}, are solutions of the equations of motion for the temperatures and  strain rates shown in the figure captions.  They all use the same system-specific physical parameters $T_P = 40,800\,K$ (implying a depinning activation energy of about 3 ev),  and $\mu/\mu_T = 31$.  Both of these parameters should be independent of strain rate and temperature. (The modulus $\mu$ by itself does depend on $T$.)  Other parameters needed for plotting these curves are $\chi_0/e_D = 0.25$ (as predicted in Sec. \ref{Teff}), plus the quantities $\kappa_{\rho}$ and $c_{e\!f\!f}$, and initial values of $\rho$ and $\chi$.  In Eq. (\ref{dotchi}) I have set $\gamma_D = 0$.

The only notable variation of these parameters from one curve to another is that the effective specific heat $c_{e\!f\!f}$ in Eq. (\ref{dotchi}) is about a factor of ten smaller for the high-temperature curves in Fig. \ref{Fig3} than it is for the lower temperature curves in Fig. \ref{Fig2}.  This difference accounts for  the sharper rise of the stress at the higher temperature.  There are no arbitrary power laws or assumptions about transitions between various ``stages'' of hardening.  The physical mechanisms that determine the shapes of these curves are contained entirely in the factors $\kappa_{\rho}$ and $c_{e\!f\!f}$. 

The conversion factor $\kappa_{\rho}$ is interesting.  Kocks and Mecking \cite{KOCKS-MECKING-03} discovered experimentally that the onset slope of these curves seemed to be a constant, independent of both strain rate and temperature -- as can be seen approximately in these figures if one adjusts for the different stress scales.  Apparently, the initial values of $\rho$ for these copper samples were much smaller than their steady-state values, so that there is no apparent yield stress, and the second term in the brackets on the right-hand side of Eq.(\ref{dotrho}) can be neglected at small strains.  Then a simple calculation tells us that

\begin{equation}
\label{onset}
{1\over \mu}\,\Bigl({\partial\sigma\over \partial\epsilon}\Bigr)_0 \cong \kappa_{\rho}\,{b^2\,\mu_T^2\,\nu_0^2\over 2\,\mu\,\gamma_D},
\end{equation}
where the subscript $0$ denotes the onset value at very small strain.  The quantity $\nu_0$ is the onset value of the slowly varying function $\nu$ given in Eq.(\ref{nudef}).  Thus, the strain rate has effectively cancelled out of this formula.  Moreover, since $\mu$, $\mu_T$, and $\gamma_D/b^2$ (each with dimensions of energy per unit volume) should all scale with temperature in about the same way, the right-hand side of Eq.(\ref{onset}) should be almost independent of temperature.  Thus, the TDT has explained the observation of Kocks and Mecking, assuming that $\kappa_{\rho}$ remains constant.

{\it Hall-Petch Effects:}  Even more interestingly, the physical interpretation of $\kappa_{\rho}$ as an energy conversion factor tells us that it cannot generally remain constant but must contain information about dislocation-creation mechanisms.  For example, the data of Meyers et al \cite{MEYERSetal-95}, as interpreted in \cite{JSL-17a}, reveals that $\kappa_{\rho}$ increases with the inverse square root of decreasing grain size.  In other words, $\kappa_{\rho}$ contains a term  proportional to the stress-concentration factor near a corner of a typical grain,  so that smaller grains are more effective sources of new dislocations than larger ones.  Then, using this conversion term to compute dislocation densities via Eq.(\ref{dotrho}) and, ultimately, yield stresses and flow stresses, we find a simple explanation of Hall-Petch grain-size effects. This explanation is far more compelling, in my opinion, than the conventional way of attributing these effects to pile-ups of dislocations at grain boundaries, and then making the manifestly incorrect assumption that the stress driving dislocations across the boundary simply adds to the stress driving them across the interior of the grain to produce the measured total stress.\cite{ARMSTRONG-HP14,ARMSTRONG-HP16}  

\begin{figure}[h]
\begin{center}
\includegraphics[width=\linewidth] {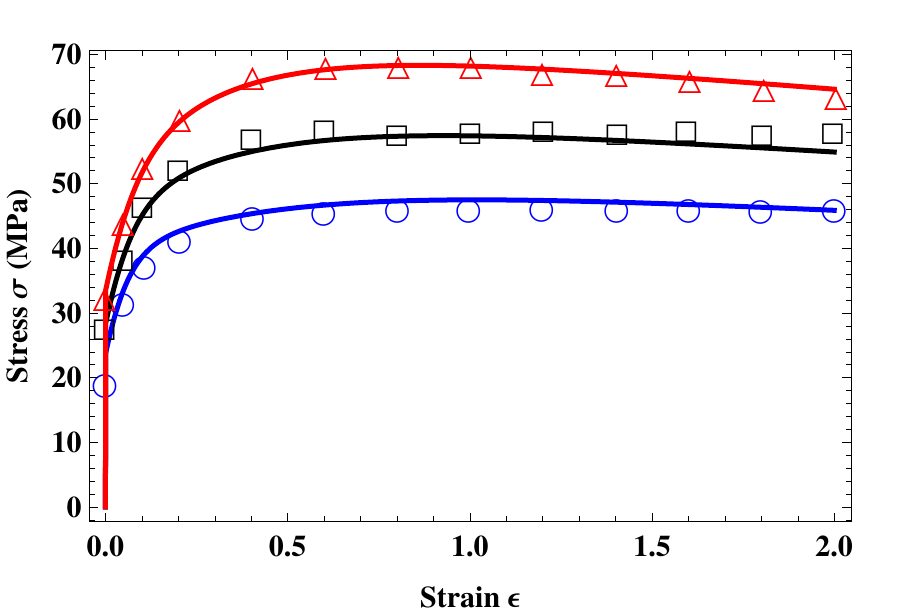}
\caption{Experimental data and theoretical stress-strain curves for aluminum at $T=573\,K$, for strain rates $0.25/s$ (lower blue curve), $2.5/s$ (middle black curve),  and $25/s$ (upper red curve). }   \label{Fig4}
 \end{center}
\end{figure}

\begin{figure}[h]
\begin{center}
\includegraphics[width=\linewidth] {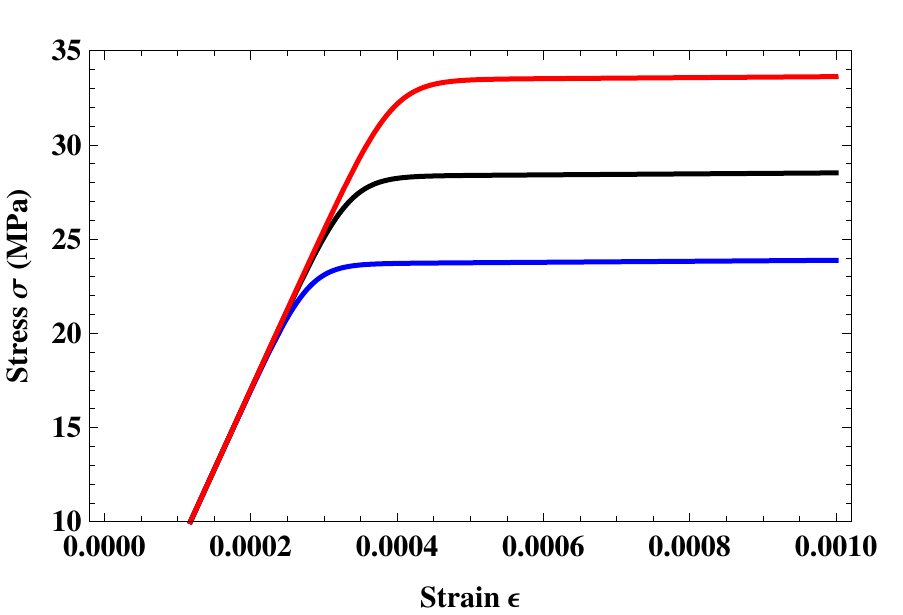}
\caption{Magnified stress-strain curves for aluminum showing the yielding transitions seen at $\epsilon \cong 0$.}   \label{Fig5}
 \end{center}
\end{figure}

{\it Strain Hardening for Aluminum:} The situation described in the preceding paragraphs for copper changes interestingly in the stress-strain curves for aluminum shown in Fig.~\ref{Fig4}. The experimental data shown here is taken from Shi et al \cite{SHIetal-97}; and the theoretical analysis is from \cite{LTL-17}.  These curves are for $T = 573\,K$, for three different strain rates $0.25/s$, $2.5/s$, and $25/s$ .    

The most obvious new feature of these curves is that they exhibit yielding transitions, near $\epsilon = 0$, at three different yield stresses for the three different strain rates.  Unlike any other dislocation theories, the TDT {\it predicts} these yielding transitions as functions of strain rate, temperature, and sample preparation.  They are not given here by the conventional von Mises or Tresca yield stresses.    

A magnified view of these three theoretical yielding transitions at strains of order $10^{-4}$ is shown in Fig.~\ref{Fig5}.  Here we see a linear elastic stress function with large slope $\mu$, rising rapidly from zero strain, and levelling off rapidly but smoothly at three different stresses determined by the three different total strain rates $\dot\epsilon$.  This behavior is governed by the extremely strong stress sensitivity  of $q(\sigma, \rho, T)$ in Eq.~(\ref{qdef}), which produces the rapid $\epsilon$-dependence of the stress in Eq.~(\ref{dotsigma}).  Note that this behavior is the converse of the extremely slow strain-rate dependence of the steady-state stress discussed at the end of Subsection \ref{Teff}.  The depinning model is playing a central role in explaining two major, ostensibly disparate features of dislocation-enabled plasticity. 

A second way in which these stress-strain curves for aluminum differ from those for copper is that, in Fig.~\ref{Fig4}, at the larger strain rate, there is clear evidence of thermal softening; the stress decreases at large strains.  To account for it, we have used Eq.~(\ref{dotT}) with parameters specified in  \cite{LTL-17}.  We shall see below that this relatively mild thermal effect plays a much more important role in the analysis of adiabatic shear banding and the onset of ductile fracture.  

{\it Adiabatic Shear Banding:} Turn now to the phenomenon of adiabatic shear banding (ASB) in which a uniformly sheared solid fails along a prescribed line, perhaps a surface scratch or a crystalline irregularity.  ``Adiabaticity''  refers to the fact that an instability is caused by thermal softening in a situation where heat flow is slower than plastic deformation. ASB is an especially useful example for our purposes because it illustrates, among other things, the strong coupling between mechanical and thermal dynamics in the TDT.  So far as I know, the TDT theory of this phenomenon is the first of its kind.

\begin{figure}[h]
\begin{center}
\includegraphics[width=\linewidth] {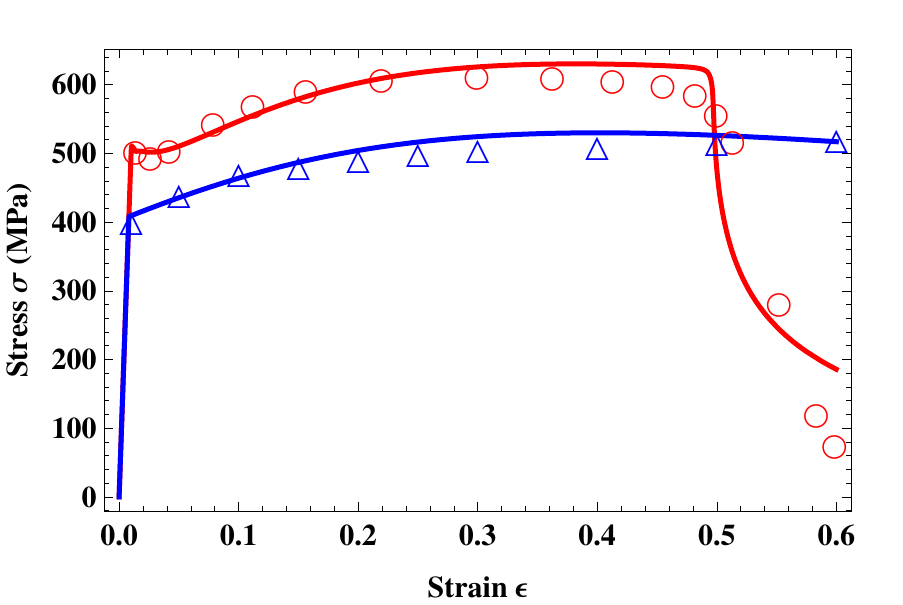}
\caption{Stress-strain curves for adiabatic shear banding in steel. The experimental data is from Fig.8 of Marchand and Duffy \cite{MARCHAND-DUFFY-88}. The upper curve (red) is for strain rate $\dot\epsilon = 3300/s$, the lower (blue) is for $10^{-4} /s$.  }   \label{Fig6}
 \end{center}
\end{figure}

The stress-strain data points  in Fig.\ref{Fig6} are taken from the classic 1988 study of ASB in steel by Marchand and Duffy.\cite{MARCHAND-DUFFY-88}   The theoretical curves are from \cite{JSL-17rev}.   See  \cite{LTL-18}  for a theoretical analysis of all the data in \cite{MARCHAND-DUFFY-88} and for graphs of space- and time-dependent strain rates and temperatures; and see \cite{JSL-17rev} for  the parameter values used for computing the particular curves shown here. 

\begin{figure}[h]
\begin{center}
\includegraphics[width=\linewidth] {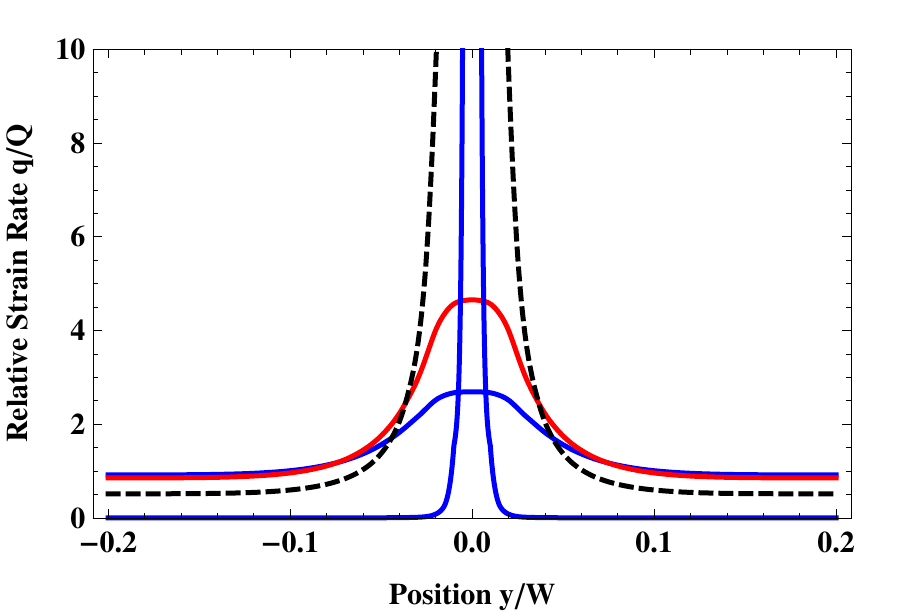}
\caption{Relative plastic strain rates $q(\epsilon,y)/Q$ at strains $\epsilon = 0.45,\,0.47,\,0.49,$ and $0.497$, for the top stress-strain curve shown in Fig.~\ref{Fig6}.  For increasing $\epsilon$, these shear flows are increasingly concentrated in a narrowing band centered at $y=0$.}   \label{Fig7}
 \end{center}
\end{figure}

The upper red stress-strain curve in Fig.\ref{Fig6} is measured at a high strain rate, $3,300/s$; the lower blue curve is effectively quasistatic, $0.0001/s$.  Both curves are measured nominally at room temperature.  By ``nominally,'' I mean that the measurements were made on samples that initially were at room temperature, but that internal heating effects were important at the high strain rate.  For both curves, $T_P = 6\times 10^5\,K$, $\mu = 5\times 10^4\,MPa$, $\mu_T= 1200\,MPa$, and $\tilde\chi_0 = 0.25$.  

Both of the stress-strain curves in Fig. \ref{Fig6} exhibit sharp yielding transitions  at small strains of order $0.02$.  These transitions are essentially the same as those shown for aluminum in Figs. \ref{Fig4} and \ref{Fig5}. The small overshoot at the yield point for the upper curve is most probably an instrumental artifact.  I reproduced it artificially here by adjusting the initial effective temperature, in effect, assuming that the overshoot was caused by sample preparation. 

The abrupt stress drop at $\epsilon \cong 0.5$ on the fast curve in Fig.\ref{Fig6} indicates the onset of the adiabatic shear-banding instability.  An increase in strain rate along the emerging shear band increases the heat generation according to the first term on the right-hand side of Eq.(\ref{dotT}).  In turn, this increase in temperature increases the local strain rate according to  Eq.({\ref{qdef}), which further increases heat generation.  The result is a runaway instability if heat is unable to flow away from the hot spot more quickly than new heat is generated there.  Thermal conduction is described by the last two terms on the right-hand side of Eq.(\ref{dotT}), which both have been set to zero in this example.  Thus, we are looking at a balance between thermal and mechanical behaviors that, in this case, is governed primarily by the strong temperature sensitivity of the depinning mechanism.  The experimentally observed stress drop is sharper than the theoretical one because  shear banding almost certainly changes into something like fracture in its late stages, and this version of the TDT is not yet a theory of fracture.  

Figure \ref{Fig7}  shows in more detail what is happening during the instability. It  shows the normalized shear rate $q(\epsilon,y)/Q$ at four different values of $\epsilon$ as the system approaches the transition.  Here, $Q/\tau_0$ is the total, externally driven strain rate, $q(y)/\tau_0 = \dot\epsilon(y)$ is the shear rate at a distance $y$ away from the band, and $W$ is the width of the sample. At first, shear localization occurs relatively slowly.  But, at about $\epsilon = 0.49$, this nonlinear process accelerates rapidly.  The plastic strain rate becomes sharply concentrated near $y=0$, causing a sudden increase in the temperature there. The stress decreases uniformly across the system, causing the strain rate to  fall toward zero everywhere except in the increasingly hot, narrowing band where the runaway instability is occurring.

\section{Livermore Molecular Dynamics Simulations}
\label{LMD}

The experimental comparisons described in the preceding Section pertain to a   range of situations in which the strain rates are much slower than the relaxation rates for the kinetic-vibrational modes, and for which the dislocation densities are  high enough that entanglements control the dynamics.  But the physical principles on which the TDT is based apply to a broader range of situations.  In fact, the original TDT paper \cite{LBL-10} contains an analysis of strong-shock measurements for which the steady-state effective temperature depended on the strain rate.   

The Livermore molecular dynamics simulations (LMD) \cite{BULATOV-17} probe an even broader range of issues than the strong-shock data.  These investigators simulated three dimensional systems containing as many as $10^{10}$ tantalum atoms interacting via realistic embedded-atom potentials. They imposed uniaxial compressive strains on brick-shaped samples with periodic boundary conditions. Using one of the world's most powerful computers, they were able to drive these systems at constant strain rates of order $10^7 /s$ and above, and reach strains of order unity.  These simulations were better than real experiments in the sense that temperatures and initial conditions were accurately controlled, there were no extraneous boundary effects, and the density of dislocations as well as the stress was measured directly. 

My analysis of the LMD simulations is described in \cite{JSL-18}. There are two important extensions of the TDT in that paper. First, the steady-state effective temperature becomes an increasing function of strain rate for $\dot\epsilon^{pl}$ above about $10^6 /s$, where the rate at which the subsystem of dislocations is being ``stirred'' becomes comparable to atomic-scale fluctuation rates.  Specifically, I wrote:
\begin{equation}
\label{q-chi}
\tilde\chi_{ss}(q) =\tilde\chi_0 + {B \over [\ln(q_0/q)]^2}.
\end{equation}
The parameter $B$ sets the crossover from constant to growing values of  $\tilde\chi_{ss}$.  The parameter $q_0$ is the dimensionless strain rate at which the subsystem of dislocations -- and thus the crystal itself -- effectively melts (as seen in the simulations).  This was my simplest, best-guess, description of this high-strain-rate effect.  So far as I can see, there is no hint of any amplified ``phonon drag'' effect here.  In the data analyses shown below, I set $B=4.6$ and $q_0 = 0.008$.

The second modification arises from the fact that the LMD simulations also explored the regime of very small dislocation densities, where the entanglement model is not relevant.  This happened at the onsets of the simulations, where the initially unstressed crystals were seeded with a few small dislocation loops.  The  increasing strain caused these loops to stretch and deform, eventually interacting with each other and with the fluctuating lattice, and creating new dislocations.  At first, the  dislocations were not dense enough to become strongly entangled; thus, the early appearance of new dislocations was a softening mechanism.  They enabled deformation, and the stress dropped.  As their density increased, they became entangled, and the system underwent conventional strain hardening.

It has been assumed since the early work of Peierls and Nabarro that a dislocation in free motion through a crystal is subject to a drag force proportional to its velocity.  Conversely, its velocity is proportional to the driving stress.  At the extremely small densities of dislocations in the initial stages of the LMD simulations, it is not reasonable to assume that the time spent by a dislocation segment in moving from one pinning site to another is neglible compared to the pinning time; thus we must account for a drag time $\tau_D$ analogous to the pinning time $\tau_P$ introduced in Eq.(\ref{tauP}), and note that these times are additive. 

The velocity $v$ appearing in the Orowan relation, Eq.(\ref{Orowan}), becomes 
\begin{equation}
v = {\ell\over \tau_P + \tau_D},
\end{equation}
where, as before, $\ell = 1/\sqrt{\rho}$.   Let $\eta$ be a drag coefficient with dimensions of stress, so that $v_{drag} \equiv \ell/ \tau_D \equiv b\,\sigma/\eta\,\tau_0$. Then, the generalization of the dimensionless strain rate $q$ defined in Eq.(\ref{qdef}) is
\begin{equation}
\label{qdef2}
q^{L\!M\!D} =\Bigl[{1\over q}+{\eta\over \sigma\,\rho} \Bigr]^{-1}.
\end{equation}
Note that the drag correction, proportional to $\eta$, requires small values of $\rho$ in order to be non-negligible.  The effective temperature plays no role in this early stage of the process. 

Note especially that the right-hand side of Eq.(\ref{qdef2}) is a highly nonlinear function of $\sigma$ and $\rho$.  It could not possibly be consistent with the conventional assumption that the stress is simply the  sum of drag and barrier-resistance terms (e.g. as in \cite{ARMSTRONG-HP14,ARMSTRONG-HP16}). But it does satisfy the condition that, when $\eta$ becomes dominantly large, $q^{L\!M\!D} \to \rho\,\sigma/\eta$.

\begin{figure}[h]
\begin{center}
\includegraphics[width=\linewidth] {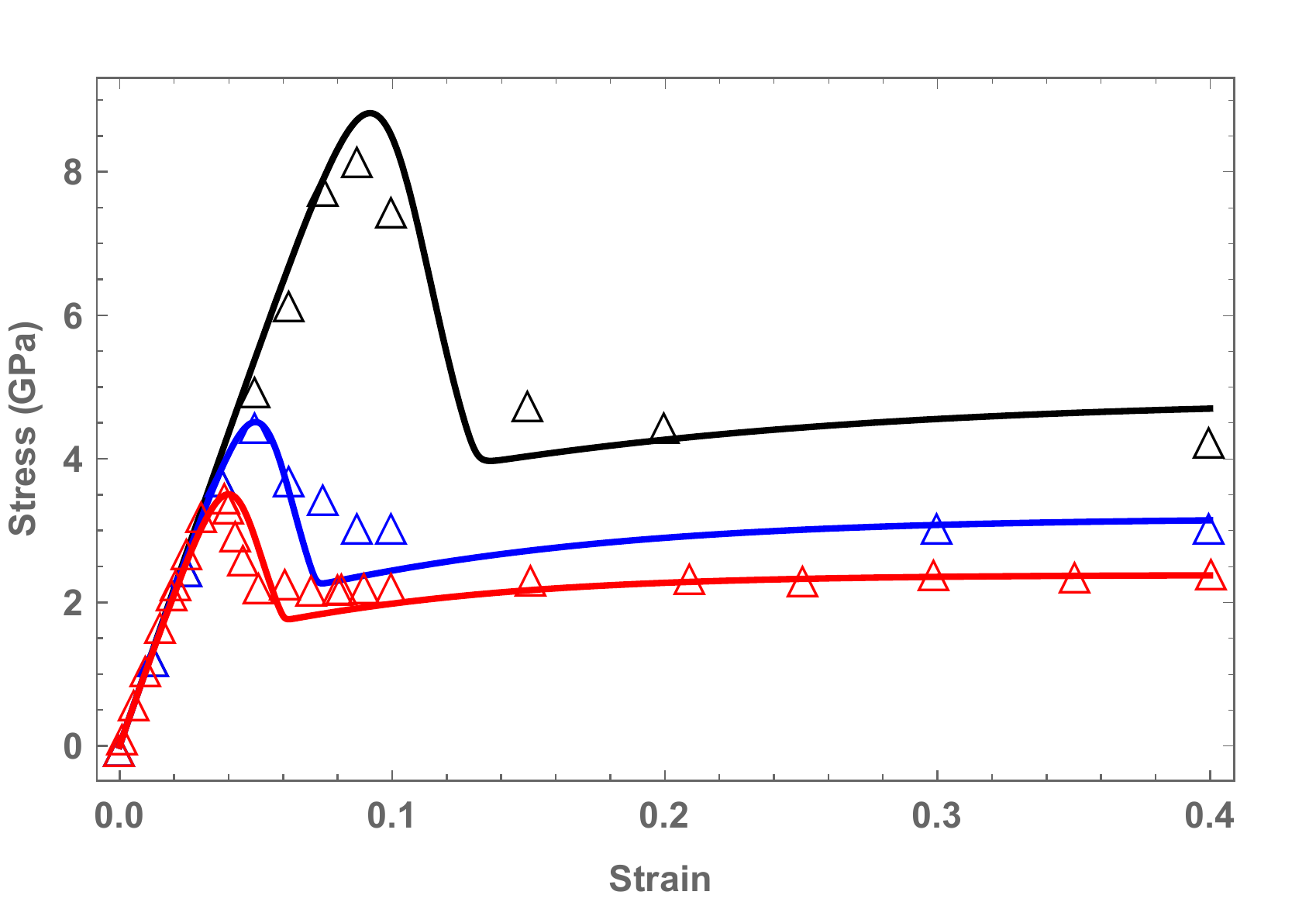}
\caption{LMD stresses as functions of strain for strain rates $\dot\epsilon = 1.1 \times 10^7 /s$, $5.55 \times 10^7/s$, and $2.77 \times 10^8 /s$, from bottom to top, plus corresponding theoretical curves.}   \label{Fig8}
 \end{center}
\end{figure}

\begin{figure}[h]
\begin{center}
\includegraphics[width=\linewidth] {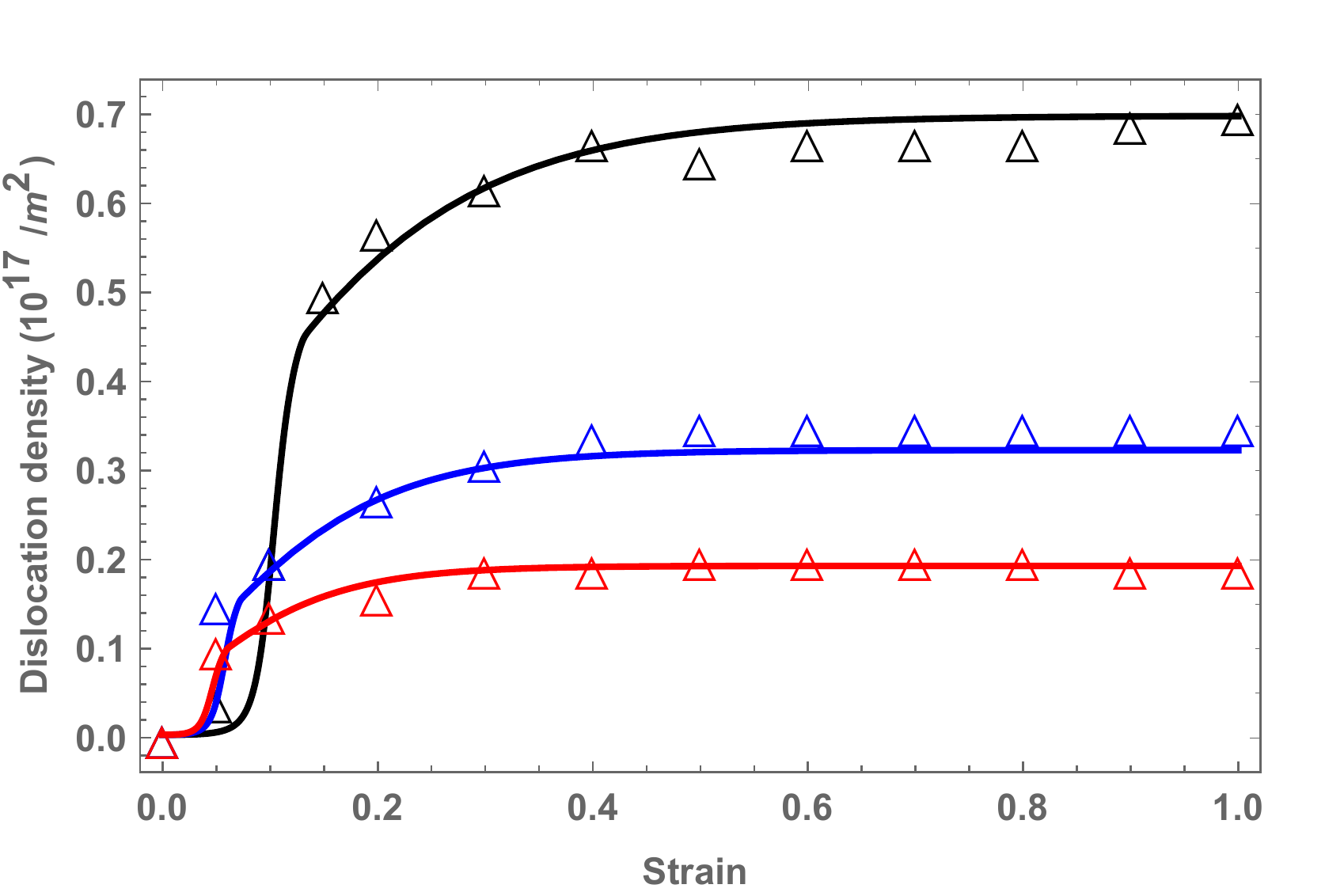}
\caption{LMD dislocation densities as functions of strain for the same strain rates shown in Fig.\ref{Fig8}, plus corresponding theoretical curves.}   \label{Fig9}
 \end{center}
\end{figure}

Both the LMD data and the theoretical results are shown in Figs. \ref{Fig8} and \ref{Fig9}.  (See \cite{JSL-18} for values of the theoretical parameters used in plotting these graphs.)  The agreement between the theory and the simulations  tells us that the extended TDT accounts satisfactorily for the crossover between drag and entanglement effects.  Note the dramatic increases in the dislocation densities in Fig.\ref{Fig9} at the points where the stresses are passing their peaks in Fig. \ref{Fig8}.  This behavior is closely related to the highly nonlinear features of the double-exponential function in Eq.(\ref{qdef}) and the highly nonlinear way in which it enters the formula for the total strain rate, Eq. (\ref{qdef2}).  Clearly, agreement between theory and simulations is good but not perfect, especially at the stress peaks.  This agreement might be improved by allowing parameters such as $\eta$ and $\kappa_{\rho}$ to be rate dependent.  There may be some physical reason why the after-peak stresses drop faster and further in the theory than in the simulations; but uncertainties in the simulation data might also be playing a role.

\section{Fracture Toughness}
\label{fracture}

This Section is different from the preceding ones.  I recently have published a TDT-based theory of fracture toughness \cite{JSL-21} that is every bit as unconventional as the theory of strain hardening described above.  However, this fracture theory is mathematically too complex and, as yet, too problematic in my opinion to be included in detail in this review.  On the other hand, I think that fracture is an overwhelmingly important topic, perhaps the single most important reason for developing a predictive theory of crystalline deformation.  Therefore, I present here just an outline of the basic ideas, without trying to show the mathematical details as comprehensively as in the earlier parts of this review, but as an attempt to describe where I think this part of the field might be going.  

Theoretical research on the physics of brittle and ductile fracture in crystalline solids has been at a decades-long standstill comparable to that which has afflicted theories of strain hardening.  Consider the following. We have long known from observation that solids are stronger when they are colder; their yield stresses and flow stresses increase with decreasing temperature.  This behavior is now  predicted by the TDT as seen in Eqs.(\ref{sigmadef}) and (\ref{nudef}).  But we also know that solids become more brittle, i.e. they break more easily at lower temperatures despite the fact that they are stronger. How can these properties be consistent with each other? 

This question has been raised by Ritchie \cite{RITCHIE-11} but, so far as I know, is not answered in the solid-mechanics literature.  The conventional model used for studying brittle or ductile crack initiation is one in which dislocations are emitted from sharp notch tips and move out along well defined slip planes.  For example, see Fig. 1 in Argon's 2001 review \cite{ARGON-01}, or the analogous description in \cite{TANAKAetal-08}.  These dislocations either move freely, supposedly implying brittle behavior, or become dense enough to shield the crack tip and somehow produce ductility and toughness.  Agreement with experiment is modest at best.  As stated in the recent experimental paper by Ast et al. \cite{ASTetal-18}, an ``understanding of the controlling deformation mechanism is still lacking.''

The experimental situation is more illuminating, but only in one case that I know about.  I refer to two remarkable papers by Gumbsch and coworkers  \cite{GUMBSCHetal-98,GUMBSCH-03} who, like Ast {\it et al.} \cite{ASTetal-18}, observed brittle and ductile behaviors of crystalline tungsten.  Gumbsch measured notch fracture-toughness as a function of temperature both below and above brittle-ductile transitions.  An example of one of his sets of measurements -- the one to which I shall pay special attention -- is shown by the triangular points in Fig. \ref{Fig10}.  

\begin{figure}[h]
\begin{center}
\includegraphics[width=\linewidth] {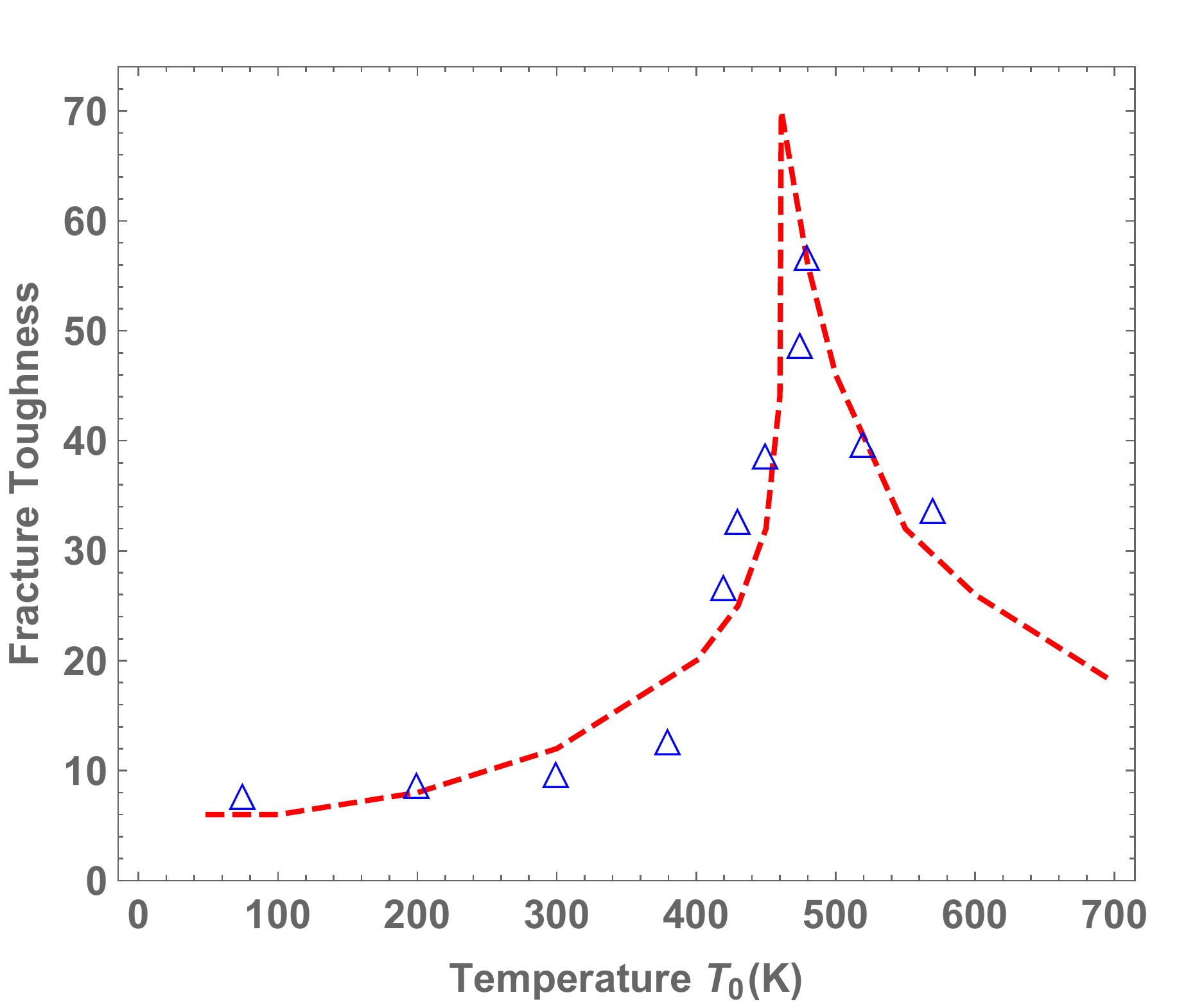}
\caption{Experimental data from \cite{GUMBSCHetal-98,GUMBSCH-03} for fracture toughness as a function of temperature (open triangles), and theoretical prediction (dashed curve), for predformed tungsten. The fracture toughness, denoted in the text by ${\cal K}_c$, is in units $M\!Pa\, m^{1/2}$; the loading rate is $0.1 M\!Pa\, m^{1/2} /s$}   \label{Fig10}
 \end{center}
\end{figure}

My analysis of the Gumbsch data is based on the TDT plus two unconventional assertions.  First: Cracks are not intrinsically sharp.  The shape (i.e. the radius of curvature) of the tip is a relevant dynamical variable. Second: Plasticity is not necessarily a blunting mechanism.  Crack initiation is controlled by the interplay between elastic and plastic dynamics near a rounded notch tip.  Especially important is the dynamics of the plastic zone at the tip, whose boundary is determined by the TDT yielding condition and, within which, the plastic strain rate is large.  If concentrated plastic deformation pulls the front of the tip forward, then the tip sharpens.  This is what we are told by both the mathematics and the experiments.  

\begin{figure}[h]
\begin{center}
\includegraphics[width=\linewidth] {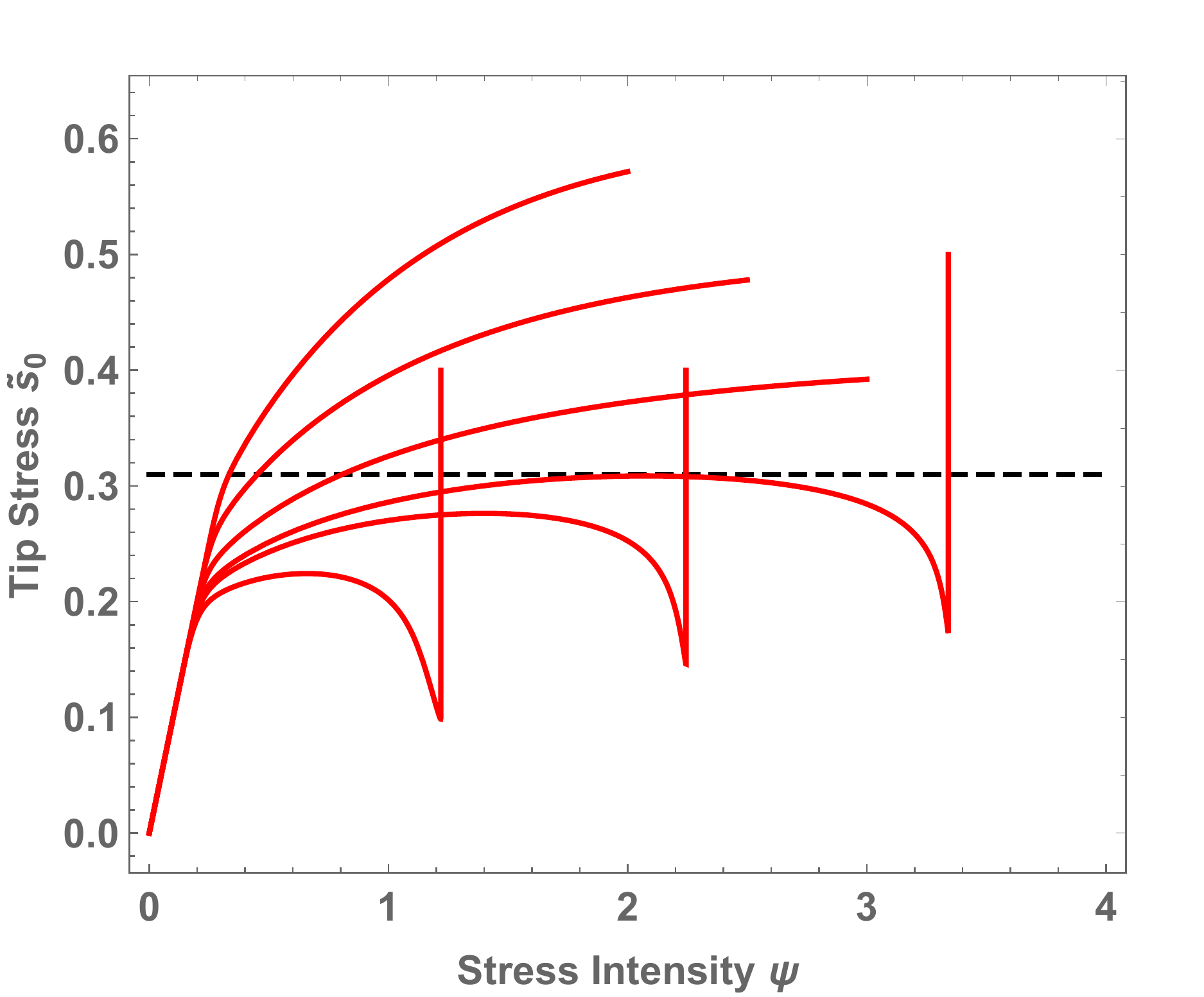}
\caption{Tip stresses $\tilde s_0(\psi)$ for temperatures $T_0 = 100 K, 200 K, 350 K, 460 K, 500 K, 600 K$ from top to bottom, plus a dashed line at the breaking stress $\tilde s_0 = \tilde s_c =0.31$. }   \label{Fig11}
 \end{center}
\end{figure}

\begin{figure}[h]
\begin{center}
\includegraphics[width=\linewidth] {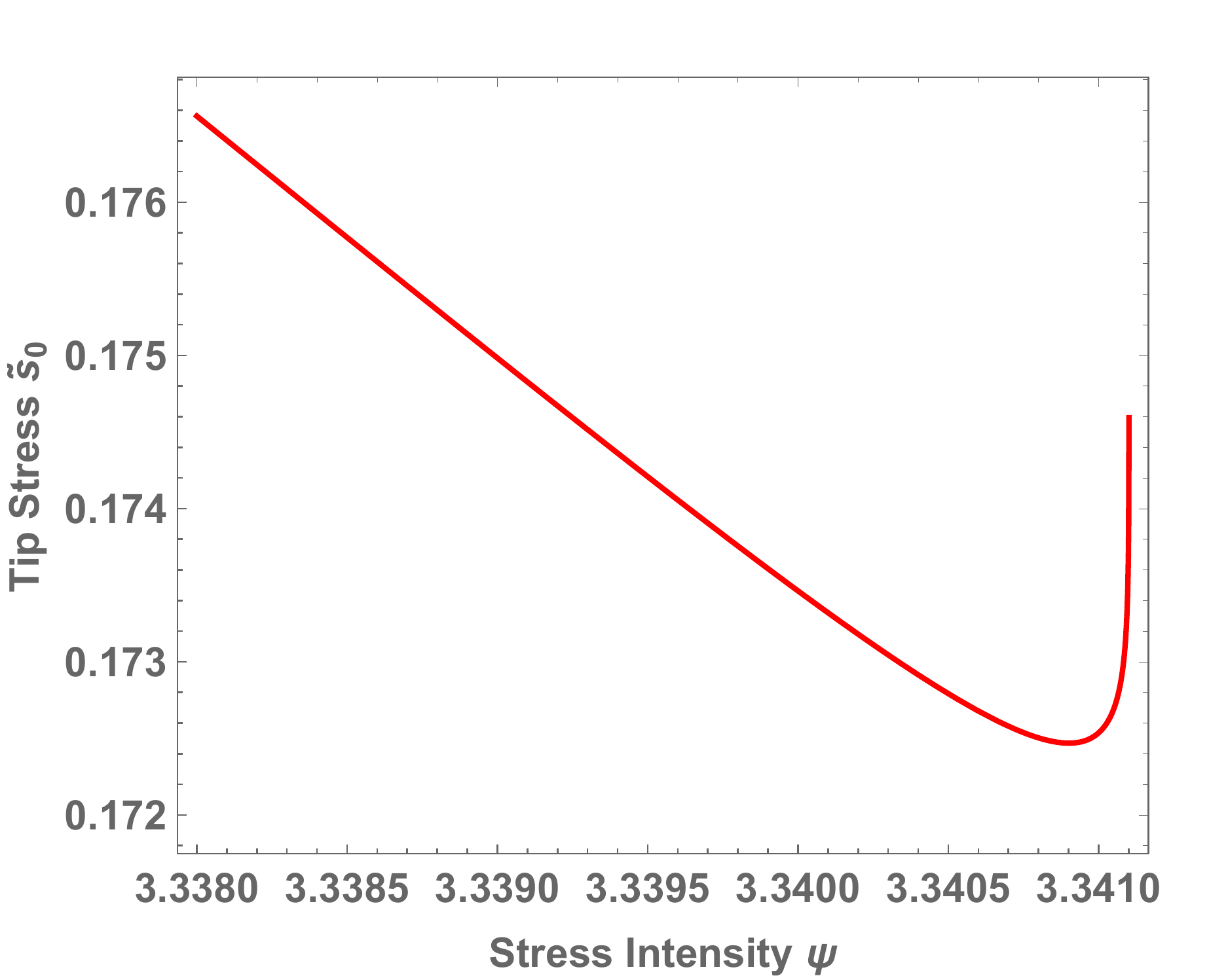}
\caption{Tip stress $\tilde s_0(\psi)$ for $T_0=460K$ near $\psi \cong 3.34$ . }   \label{Fig12}
 \end{center}
\end{figure}

The analysis in \cite{JSL-21}, like that in my earlier theory of fracture in metallic glasses \cite{JSL-20},(see also \cite{SCHetal-13,SCHetal-18,RB-12}) starts by describing this notch as the front half of an elongated elliptical hole oriented along the $x$ axis, in a material driven at infinity by a mode-1 stress $\sigma_\infty$ in the $y$ direction. Let the initial length of this elliptical notch be $W$, and let the initial radius of curvature at its front end be $d_{t\!i\!p} \ll W$.  Then define a dimensionless stress-intensity factor
\begin{equation}
\psi \equiv {\sigma_{\infty}\over \mu_T} \sqrt{2 W\over d_{t\!i\!p}}
\end{equation}
where $\mu_T$ is the reduced shear modulus defined in Eq.(\ref{sigmaT}). If, as in the experiments, the driving stress $\sigma_{\infty}$ is increased at a constant rate, then we can use $\psi$ as a dimensionless, time-like variable.  

The data shown in Fig. \ref{Fig10} is for a ``pre-hardenened'' crystal, that is, one   that has undergone strain hardening and therefore contains a high density of dislocations.  This is important because, as noted following Eq.(\ref{Orowan}), the Orowan relation between strain rate and dislocation density would not be accurate in a situation where the spacing between dislocations is larger than or comparable to the size of the notch tip.  For this pre-hardened case, I assume that the Orowan relation is correct. 

In \cite{JSL-21}, I derived an approximate set of coupled nonlinear equations of motion for the deviatoric stress at the notch tip, $\tilde s_0$ (in units of the reduced shear modulus $\mu_T$), the curvature of the tip, the dislocation density at the tip, and the temperature $T$ at the tip, all as functions of the steadily increasing, dimensionless stress-intensity factor $\psi$.  (I did not include a variable effective temperature.) Solutions of these equations for the tip stress $\tilde s_0(\psi)$, for six selected starting temperatures $T_0$, are shown in Fig.\ref{Fig11}.  These curves describe most of the important physics.

First, note that all six curves in Fig.\ref{Fig11} start on a single straight line $\tilde s_0 =\psi$, which is the purely elastic concentrated stress at the as-yet unshielded notch tip. These curves rise linearly and then bend forward at temperature-dependent  stresses indicating the formation of a plastic zone that shields the tip. These plastic yield points are analogous to those shown in Fig.\ref{Fig5} for the onset of hardening in aluminum, except that these are determined by changes in temperature instead of strain rate.  They are higher at the lower temperatures as predicted in Eqs.(\ref{sigmadef}) and (\ref{nudef}).

The most dramatic features of the curves in Fig.\ref{Fig11} are the sharp upward reversals where the tip stresses rise suddenly as the plastic zone breaks down, indicating the onset of large-scale ductile failure.  An expanded picture of this divergence of $\tilde s_0(\psi)$ is shown in Fig.\ref{Fig12}, making it clear that this rapid upturn is mathematically smooth.  This abrupt divergence of the stress is accompanied by a divergence of the temperature $T$, roughly analogous to the thermal spike that accompanies the onset of adiabatic shear banding.   

The last assumption needed in order to compute the theoretical values of the fracture toughness ${\cal K}_c$ shown in  Fig.\ref{Fig10} is that brittle fracture is initiated when the dimensionless deviatoric tip stress $\tilde s_0(\psi)$ reaches a critical value, say $\tilde s_c $, indicated by the horizontal dashed line in Fig.\ref{Fig11}.  The values of ${\cal K}_c$ are proportional to the values of $\psi = \psi_c$ at which the stress curves cross ${\tilde s}_c$, that is, $\tilde s_0(\psi_c) = {\tilde s}_c$.  I estimate that $\tilde s_c\cong 0.31$ and that ${\cal K}_c \cong 20\,\psi_c$ for this example.  

The theoretical fracture-toughness curve shown in Fig. \ref{Fig10} was constructed by computing $\tilde s_0(\psi_c)$ at 14 different initial temperatures $T_0$. At the lowest temperatures, plastic shielding is negligable and the fracture toughness levels off at $\psi_c \approx \tilde{s}_c$, i.e., at  ${\cal K}_c \cong 6$.  At the highest initial temperatures, the shielding effect is strong and the tip stress does not reach $\tilde{s}_c$ before the boundary layer undergoes its thermal instability indicated by the (nearly) vertical lines in the figure.  Those intersections automatically give us values of $\psi_c$ on the ductile part of the curve.   The one exception is the $\tilde s_0(\psi)$ curve at $T_0 = 460\,K$ whose peak is tangent to the horizontal line at $\tilde{s}_c = 0.31$.  Here I assume that the toughness jumps discontinuously from the tangent point to the thermal instability point, and that this is what is interpreted experimentally as the brittle-ductile transition.  Surely this is only an approximation.  In any case, Ritchie's conflicts \cite{RITCHIE-11} seem to be resolved.

On the whole, given the uncertainties in both the theory and the experiments, this nontrivial agreement with experiment over a temperature range of $500\,K$ leads me to believe that this part of the theoretical picture is basically correct.  But the picture is less clear for the non-predeformed crystals where the initial dislocation densities are too small to be consistent with the Orowan formula. In \cite{JSL-21}, I dealt with this situation by introducing a separate family of dilute, non-entangled dislocations with which I was able to fit Gumbsch's data.  But more work is needed -- both theoretical and experimental -- to correct and complete this analysis.

\section{Questions}
\label{next}

At the beginning of Sec.\ref{TDT}, I remarked that the version of the TDT presented here is a ``caricature'' of a realistic dislocation theory. The body of this review has been devoted to showing that this caricature captures many of the known central features of dislocation dynamics that are missing in the conventional phenomenological approaches.  

Obviously, we need to find out what happens when we restore realistic details to the TDT. A good start has been made by K.C. Le \cite{Le-18,Le-19}, who has used a combination of the TDT and conventional analyses to study nonuniform deformations, torsions and the like. However, neither of us has yet to address some fundamental questions.  Here are a few examples. 

Under what circumstances, and for what purposes, do we need to make distinctions between dislocations moving on different slip planes?  In principle, it should not be difficult to make such distinctions by describing different populations of dislocations by different order parameters, i.e. by different densities $\rho$, and describing how they transform to one another.  What will we learn by doing that?  Will we see different stages of strain hardening? Lattice rotations? Or other phenomena that are missing in TDT? Or will we see just quantitative corrections to qualitatively correct behaviors already described by this theory? 

Similarly, how will TDT explain the differences in mechanical behaviors between fcc, bcc, and hcp crystals? Will we need only to change the values of material parameters such as $\kappa_{\rho}$ and $c_{e\!f\!f}$ in Eqs. (\ref{dotrho}) and (\ref{dotchi})?  Or will more substantial changes be needed, such as those suggested in the preceding paragraph?

What about the elastic interactions between dislocations?  Presumably these are responsible for producing the cellular dislocation patterns visible in many micrographs.  Are these patterns related to rates of strain hardening?  Are they causes or effects?  I have argued that they are secondary effects that occur only at relatively small strain rates; but I certainly may be wrong. (See the end of Sec.II in \cite{JSL-16}.) What other phenomena might be related to the elastic interactions?

How do we understand Bauschinger effects, i.e. the asymmetries often seen in stress reversals?  Is Le's explanation in \cite{Le-19} sufficient?  Or must we look harder for something like irreversible lattice deformations?  

How might the TDT predict the enhanced strain hardening observed in some alloys? This should be an interesting theoretical problem because diffusion times for solute atoms may be comparable to depinning times for dislocations, and localized compositional variations may modify pinning strengths.   

A possibly  related question is whether the TDT can be used to understand the Portevin-Le-Chatelier (PLC) effect, where the interplay between dislocation dynamics and solute diffusion produces repeated shear banding instabilities at very small driving rates.  \cite{ANANTH}  This brings me back to my question at the beginning of subsection \ref{Teff} about how the TDT relates to nearly deterministic regimes in which the effective temperature may be irrelevant. 

These are short-term questions.  My asking them here means that I think they can be answered, perhaps not easily, but within the basic concepts of the TDT.  On the other hand, there is a more fundamental issue that needs to be addressed.  

In my introduction to the TDT, in subsection \ref{Teff} of this review, I pointed out that the dislocations constitute a distinct subset of the degrees of freedom of a deforming crystalline solid.  We can compute the energy and the entropy of this subsystem with little or no ambiguity; thus we can compute its effective temperature, and move on systematically from there. Are there any other realistic systems for which a comparable separability is accurately true? How broadly valid is the effective-temperature concept? 

The idea of the effective temperature arose (for me) in the study of amorphous solids where the flow defects -- the analogs of the dislocations -- are shear transformation zones (STZ's).  These are ephemeral, small-scale, structural irregularities whose reorientations produce localized plastic deformations.  They are qualitatively unlike the highly visible dislocations. We cannot look at a deforming amorphous material and predict where the next STZ is going to appear.   We can make statistical models of these processes using effective-temperature ideas \cite{FL-11,JSL-15 Yielding,H-L-07}; and those models have been useful for describing yielding behaviors and even fracture \cite{RB-12,JSL-20}.  But those models, so far, have not successfully described glass transitions or viscosity singularities in realistic, finite-dimensional systems with finite-ranged interactions, and most likely cannot do so.  The STZ analysis is not so reliably predictive as the dislocation theory seems to be.  

Turbulence is yet another story.  Contrary to Cottrell's opinion, turbulence looks today like a far more difficult problem than strain hardening.  We have known for decades about cascades of turbulent fluctuations, ``puffs'' in pipe flows, etc.; and we are learning more than we may like to know about turbulent instabilities in our attempts to predict the weather and climate change.  But I see no hint of there being a unifying concept analogous to the effective temperature.  I may very well be wrong.  I hope so. 

In short, dislocation theory may be turning out to be narrowly special; but that is not an excuse for ignoring it.  The strength and toughness of crystalline solids is too broadly important a research area for it to have been neglected by theoretical physicists for so long. We need now to bring it up to a predictive, scientific level comparable to that of other parts of materials and condensed matter physics.


\begin{thebibliography}{99}

\bibitem{LBL-10} J.S. Langer, E. Bouchbinder and T. Lookman, Thermodynamic theory of dislocation-mediated plasticity, Acta Mat. {\bf 58}, 3718 (2010).

\bibitem{TAYLOR-34} G.I. Taylor, The mechanism of plastic deformation in crystals, Part 1 - Theoretical,  Proc. Roy. Soc. A 145, 362 (1934).

\bibitem{COTTRELL-53} A.H. Cottrell, {\it Dislocations and Plastic Flow in Crystals}, (Oxford University Press, London, 1953).

\bibitem{FRIEDEL-67} J. Friedel, {\it Dislocations} (Pergamon, Oxford, 1967).

\bibitem{HIRTH-LOTHE-68} J. Hirth and J. Lothe, {\it Theory of Dislocations}, (McGraw Hill, New York, 1968).

\bibitem{COTTRELL-02} A.H. Cottrell, in {\it Dislocations in Solids}, vol. 11, F.R.N. Nabarro, M.S. Duesbery, Eds. (Elsevier, Amsterdam, 2002), p. vii. 

\bibitem{GRAY-12} G.T. Gray III, High-strain-rate deformation:
Mechanical behavior and deformation substructures induced, Annu. Rev. Mater. Res. {\bf 42}, 285 (2012).

\bibitem{ARMSTRONG-HP14} R.W. Armstrong, 60 Years of Hall-Petch: Past to present nano-scale connections, Materials Transactions {\bf 55}, 2-12 (2014). (Special issue on strength of fine grained materials, The Japan Institute of Metals and Materials, 2013).

\bibitem{ARMSTRONG-HP16} R.W. Armstrong, Hall-Petch description of nanopolycrystalline Cu, Ni and Al strength levels and strain rate sensitivities, Phil. Mag. {\bf 96} , 3097 (2016).

\bibitem{Barton etal-2011} N. Barton et al, A multiscale strength model for extreme loading conditions, J. Applied Phys. 109(7), 073501 (2011).

\bibitem{LESARetal-06} Z. Wang , N. Ghoniem, S. Swaminarayan, and R. LeSar, A parallel algorithm for 3D dislocation dynamics, J. Computational Phys. {\bf 219}, 608 (2006) .

\bibitem{KUBIN-08} B. Devincre, T. Hoc and L. Kubin, Dislocation mean free paths and strain hardening of crystals, Science {\bf 320}, 1745 (2008).

\bibitem{LESAR-14} R. LeSar, Simulations of dislocation structure and response, Annu. Rev. Cond. Matter Phys. {\bf 5}, 375 (2014).

\bibitem{SILLS-18} R.B. Sills, N. Bertin, A. Aghaei, and W. Cai, Dislocation networks and the microstructural origin of strain hardening, Phys. Rev. Lett. {\bf 121}, 085501 (2018).

\bibitem{BULATOV-17} L.A. Zepeda-Ruiz, A. Stukowski, T. Oppelstrup and V.V. Bulatov, Probing the limits of metal plasticity with molecular dynamics simulations, Nature 23472  (2017).

\bibitem{JSL-17} J.S. Langer, Thermal effects in dislocation theory. II. Shear banding, Phys. Rev. E {\bf 95}, 013004 (2017).

\bibitem{JSL-17a} J.S. Langer, Yielding transitions and grain-size effects in dislocation theory, Phys. Rev. E {\bf 95}, 033004 (2017).

\bibitem{JSL-17rev} J.S. Langer, Thermodynamic theory of dislocation-enabled plasticity, Phys. Rev. E {\bf 96}, 053005 (2017).

\bibitem{LTL-17} K.C. Le, T.M.Tran and J.S. Langer, Thermodynamic dislocation theory of high-temperature deformation in aluminum and steel, Phys. Rev. E {\bf 96}, 013004 (2017).

\bibitem{LTL-18} K.C. Le, T.M.Tran and J.S. Langer, Thermodynamic dislocation theory of adiabatic shear banding in steel, Scripta Mat.  {\bf 149}, 62 (2018) .

\bibitem{JSL-18} J.S. Langer, Thermodynamic analysis of the Livermore molecular-dynamics simulations of dislocation-mediated plasticity, Phys. Rev. E {\bf 98}, 023006 (2018). 

\bibitem{JSL-PCH-19} J.S. Langer, Statistical thermodynamics of crystal plasticity, Journal of Statistical Physics, {\bf 175}, 531 (2019).

\bibitem{FL-11} M. L. Falk and J. S. Langer, Deformation and failure of amorphous, solidlike materials, Annu. Rev. Condens. Matter Phys. {\bf 2}, 353 (2011). 

\bibitem{PTW-03} D.L. Preston, D.L. Tonks, and D.C. Wallace, Model of plastic deformation for extreme loading conditions, J. Appl. Phys. {\bf 93}, 211 (2003). 

\bibitem{KCL-20} K.C. Le, Two universal laws for plastic flows and the consistent thermodynamic dislocation theory, Mech. Res. Commun. {\bf 109}, 103597 (2020). 

\bibitem{JSL-KCL-20} J.S. Langer and K.C. Le, Scaling confirmation of the thermodynamic dislocation theory, Proc. Nat. Acad. Sci. {\bf 117} 29431 (2020).

\bibitem{SAMANTA} S.K. Samanta, Dynamic deformation of aluminum and copper at elevated temperatures, J. Mech. Phys. Solids {\bf 19} 117-135 (1971).

\bibitem{LANL-99} S.R. Chen, P.J. Maudlin, and G.T. Gray, III,  {\it Seventh International Symposium on Plasticity and Its Current Applications}, pp. 623-626, A.S. Khan, ed. (Cancun, Mexico, Neat Press, 1999).

\bibitem{KOCKS-MECKING-03} U.F. Kocks and H. Mecking, Physics and phenomenology of strain hardening: the FCC case, Prog. Matls. Sci. {\bf 48}, 171 (2003).

\bibitem{MEYERSetal-95} M. Meyers, U. Andrade and A. Chokshi, The effect of grain size on the high-strain, high-strain-rate behavior of copper, Metall. and Materials Trans. A {\bf 26A}, 2881 (1995).

\bibitem{SHIetal-97} H. Shi, A.J. McLaren, C.M. Sellars, R. Shahani, and R. Bolingbroke, Mater. Sci. Tech. {\bf 13}, 210 (1997).

\bibitem{MARCHAND-DUFFY-88} A. Marchand and J. Duffy, An experimental study of the formation process of adiabatic shear bands in a structural steel, J. Mech. Phys. Solids {\bf 36}, 251 (1988).

\bibitem{JSL-21} J.S. Langer, Fracture toughness of crystalline solids, Phys. Rev. E {\bf 103}, 063004 (2021).

\bibitem{RITCHIE-11} R.O. Ritchie, The conflicts between strength and toughness, Nat. Mater. {\bf 10}, 817 (2011).  

\bibitem{ARGON-01} A.S. Argon, Mechanics and Physics of Brittle to Ductile Transitions in Fracture, Journal of Engineering Materials and Technology {\bf 123}, 1 (2001).

\bibitem{TANAKAetal-08} M. Tanaka, E. Tarleton and S.G. Roberts, The brittle–ductile transition in single-crystal iron, Acta Mat. {\bf 56} 5123 (2008).

\bibitem{ASTetal-18} J. Ast, J.J. Schwiedrzika, J. Wehrs, D. Frey, M.N. Polyakov, J. Michler, X. Maeder, The brittle-ductile transition of tungsten single crystals at the micro-scale, Materials and Design {\bf 152}, 168 (2018). 

\bibitem{GUMBSCHetal-98} P. Gumbsch, J. Riedle, A. Hartmaier and H. F. Fischmeister, Controlling factors for the brittle-to-ductile transition in tungsten single crystals, Science {\bf 282}, 1293 (1998).

\bibitem{GUMBSCH-03} P. Gumbsch, Brittle fracture and the brittle-to-ductile transition of tungsten, J. Nuclear Materials {\bf 323}, 304 (2003).

\bibitem{JSL-20} J.S. Langer, Brittle-ductile transitions in a metallic glass,  Phys. Rev. E {\bf 101}, 063004 (2020).

\bibitem{SCHetal-13} G. Kumar, P. Neibecker, Y. Liu and J. Schroers, Nature Communications {\bf 4}, 1536 (2013).

\bibitem{SCHetal-18} J. Ketkaew, W. Chen, H. Wang, A. Datye, M. Fan, G. Pereira, U. Schwartz, Z. Liu, R. Yamada, W. Dmowski, M. Shattuck, C. O'Hern, T. Egami,  E. Bouchbinder, and J. Schroers, Mechanical glass transition revealed by the fracture toughness of metallic glasses, Nature Communications {\bf 9}, 3271 (2018).

\bibitem{RB-12} C.H. Rycroft and E. Bouchbinder, Fracture toughness of metallic glasses: Annealing-induced embrittlement, Phys. Rev. Lett. {\bf 109}, 194301 (2012).


\bibitem{Le-18} K.C. Le, Thermodynamic dislocation theory for nonuniform plastic deformations, J. Mech. Phys. Solids {\bf 111}, 157 (2018).

\bibitem{Le-19} K.C. Le, Thermodynamic dislocation theory: Finite deformations, International Journal of Engineering Science {\bf 139}, 1 (2019).

\bibitem{JSL-16} J.S. Langer, Thermal effects in dislocation theory, Phys. Rev. E {\bf 94}, 063004 (2016).

\bibitem{ANANTH} G. Ananthakrishna, Current theoretical approaches to collective behavior of dislocations,  Phys. Rep. {\bf 440}, 113-259 (2007).

\bibitem{JSL-15 Yielding} J.S. Langer, Shear-tranformation-zone theory of yielding in athermal amorphous materials, Phys. Rev. E {\bf 92}, 012318 (2015).

\bibitem{H-L-07}T.K. Haxton and A.J. Liu, Activated dynamics and effective temperature in a steady state sheared glass, Phys. Rev. Lett. {\bf 99}, 195701 (2007). 




\end{thebibliography}
\end{document}